\documentclass[aps,pra,superscriptaddress,twocolumn,10pt]{revtex4-2}
\usepackage{graphicx}
\usepackage{dcolumn}
\usepackage{bm}
\usepackage{nicefrac}
\usepackage{amsfonts,amsmath,amssymb,stmaryrd}
\usepackage{braket}
\usepackage{tabularx}
\usepackage{multirow} 
\usepackage{hhline}
\usepackage{subfigure}  
\usepackage{bbm} 
\usepackage[pdftex]{epsfig}
\usepackage{mathrsfs}
\usepackage{verbatim}
\usepackage{ulem}
\usepackage{array}
\usepackage{cancel}
\usepackage{ifthen}
\usepackage{float} 
\usepackage{listings}
\usepackage{color}
\usepackage{hyperref}
\usepackage{amsmath}
\hypersetup{colorlinks=true,linkcolor=blue,anchorcolor=blue,citecolor=blue,filecolor=blue,urlcolor=blue}
\let\revappendix\appendix
\usepackage[nameinlink,noabbrev]{cleveref}

\begin{document}

\title{Measurement of Casimir-Polder interaction for slow atoms through a material grating}

%%%%%%%%%%%%%%%%%%%%%%%%%%%%%%%%%%%%%%%%%%%%%
\author{Julien Lecoffre}
\affiliation{Laboratoire de Physique des Lasers, Université Sorbonne Paris Nord, CNRS UMR 7538, F-93430, Villetaneuse, France.}

\author{Ayoub Hadi}
\affiliation{Laboratoire de Physique des Lasers, Université Sorbonne Paris Nord, CNRS UMR 7538, F-93430, Villetaneuse, France.}

\author{Matthieu Bruneau}
\affiliation{Laboratoire de Physique des Lasers, Université Sorbonne Paris Nord, CNRS UMR 7538, F-93430, Villetaneuse, France.}
\affiliation{Leibniz University of Hanover, Institute of Quantum Optics, QUEST-Leibniz
Research School, Hanover, Germany.}

\author{Charles Garcion}
\affiliation{Laboratoire de Physique des Lasers, Université Sorbonne Paris Nord, CNRS UMR 7538, F-93430, Villetaneuse, France.}
\affiliation{Leibniz University of Hanover, Institute of Quantum Optics, QUEST-Leibniz
Research School, Hanover, Germany.}

\author{Nathalie Fabre}
\affiliation{Laboratoire de Physique des Lasers, Université Sorbonne Paris Nord, CNRS UMR 7538, F-93430, Villetaneuse, France.}

\author{Éric Charron}
\affiliation{Université Paris-Saclay, CNRS, Institut des Sciences Moléculaires d’Orsay, F-91405 Orsay, France.}

\author{Naceur Gaaloul}
\affiliation{Leibniz University of Hanover, Institute of Quantum Optics, QUEST-Leibniz Research School, Hanover, Germany.}

\author{Gabriel Dutier}
\affiliation{Laboratoire de Physique des Lasers, Université Sorbonne Paris Nord, CNRS UMR 7538, F-93430, Villetaneuse, France.}

\author{Quentin Bouton}
\affiliation{Laboratoire de Physique des Lasers, Université Sorbonne Paris Nord, CNRS UMR 7538, F-93430, Villetaneuse, France.}

%%%%%%%%%%%%%%%%%%%%%%%%%%%%%%%%%%%%%%%%%%%
\date{\today}
%%%%%%%%%%%%%%%%%%%%%%%%%%%%%%%%%%%%%%%%%%%
\begin{abstract}
We present a method utilizing atomic diffraction patterns and statistical analysis tools to infer the Casimir-Polder interaction between Argon atoms and a silicon nitride nanograting. The quantum model that supports the data is investigated in detail, as are the roles of nanograting geometry, finite size effects, slit width opening angles, and Lennard-Jones potentials. Our findings indicate that the atom-surface potential strength parameter is $C_{3} = 6.87 \pm 1.18$\,meV.nm$^3$. This value is primarily constrained by the knowledge of the nanograting geometry. The high sensitivity of our method paves the way for precise determination of the Casimir-Polder potential and exploration of new short-distance forces. 
\end{abstract}
%%%%%%%%%%%%%%%%%%%%%%%%%%%%%%%%%%%%%%%%%%%
\maketitle

\section{\label{sec:level1}Introduction}

The Casimir-Polder (C-P) force represents one of the most fascinating manifestations of vacuum energy fluctuations. This electromagnetic force arises between an atom and a surface, and it is a consequence of matter polarization induced by atomic electromagnetic field fluctuations \cite{Buhmann2012}. The associated interaction potential encompasses a variety of dispersion forces, including those dominated by electrostatic effects at short distances, which are known as van der Waals forces, and those influenced by retardation effects due to the finite speed of light.

This force offers the potential to engineer interactions between atoms, light, and macroscopic objects. Consequently, the range of quantum technologies involving atoms positioned near surfaces has significantly broadened, encompassing nanofibers \cite{Mitsch2014,Vetsch2010, Deasy2014, Patterson2018}, nanocells \cite{Peyrot2019, Peyrot2019_v2}, waveguides \cite{Ritter2018, Skljarow2022}, and atom chips \cite{Knappe2004,Nshii2013,Amico2017}. As these technologies continue to be miniaturized at an accelerated pace, many of these systems are reaching a scale where the C-P force is becoming increasingly important for the large-scale deployment of quantum technologies. In this context, a deep understanding of the interactions between atoms and surfaces is of paramount importance. Moreover, an accurate measurement and description of the atom-surface interaction is also essential for investigating hypothetical new short-distance forces. From the perspective of fundamental science, numerous theories propose modifications to the Newtonian gravitational interaction at the nanoscopic scale. Experimental confirmation of these predictions would be of great significance. However, the preponderance of the C-P force at this scale represents a significant experimental challenge for probing these hypothetical forces \cite{Onofrio2006,Decca2007}. Consequently, C-P interaction measurements are of great importance. Yet, experimental measurements remain quite rare, and the development of a precise methodology to accurately extract information on the C-P potential parameters is an ongoing quest.

Inferring the C-P force is a challenging task, as it requires precise control of the atom-surface distance in a given internal state \cite{Laliotis2021}. Over the past decade, a variety of C-P force measurements have been demonstrated, employing a range of techniques. These include the use of ultra-cold atoms \cite{Bender2014, Balland2023}, spectroscopy methods with nanoscopic cells \cite{Peyrot2019_v2, Whittaker2014, Carvalho2018} and atomic diffraction through material nanogratings using a supersonic \cite{Lonij2009} or slow atomic beam \cite{Garcion2021, Morley2021}. Despite the advancement of these techniques, challenges remain. These include the presence of electrostatic forces from adsorbed atoms, difficulties in estimating collisional processes in dense vapors, and theoretical challenges in describing the system dynamics. While these techniques could yield the C-P potential strength parameter $C_{3}$, to the best of our knowledge, none of these experiments have inferred this $C_{3}$ parameter through a metrological study with an accepted statistical model that accounts for systematic effects, such as the finite size and opening angles of the gratings.

In this study, we investigate the C-P interaction using atomic diffraction of noble Argon gas with a nanograting. The experiment employs atomic beams with velocities below 16\,m/s, resulting in diffraction patterns that are predominantly influenced by atom-surface interactions. This approach overcomes significant limitations faced in previous experiments. Specifically, the use of Argon atoms prevents atomic adsorption on the nanograting, while the low density of the atomic beam prevents collision effects. To conclude our measurements, we extract information about the atom-surface interaction through a comprehensive systematics study using advanced statistical tools and a quantum model that we have recently developed \cite{Garcion2024}. While previous studies have mainly focused on demonstrating the possibility of determining the $C_{3}$ parameter, our study aims to rigorously validate the theoretical model, thus enabling a precise investigation of statistical and systematic effects on the measurements. This procedure allows us to bound and characterize the C-P interaction with an accuracy of 17.2\,\%. This approach enables precise measurement of the atom-surface interaction potential, thereby paving the way for the full sensitivity and characterization of the C-P potential and the exploration of new forces at short range.

The structure of the paper is as follows: In Section~\ref{sec:level2}, we describe the experimental set-up. Section~\ref{sec:level3} briefly presents the theoretical model. In Section~\ref{sec:level4}, the experimental data are subjected to statistical analysis in order to assess their compatibility with the theoretical results. In this section, we demonstrate a statistical error of 1\,\% at the 95\,\% confidence interval, which is limited only by the quantum shot noise. This section also considers the effects of the atomic source and of the Lennard-Jones potential. In Section~\ref{sec:level5}, we concentrate our analysis on systematic errors, such as the nanograting geometric parameters, the influence of the finite size of the grating and the opening angle of the slits. Finally, a summary and conclusion are presented in Section~\ref{sec:level6}.

\section{\label{sec:level2}Experimental set-up}

\begin{figure}[t]
{\includegraphics[width=\linewidth]{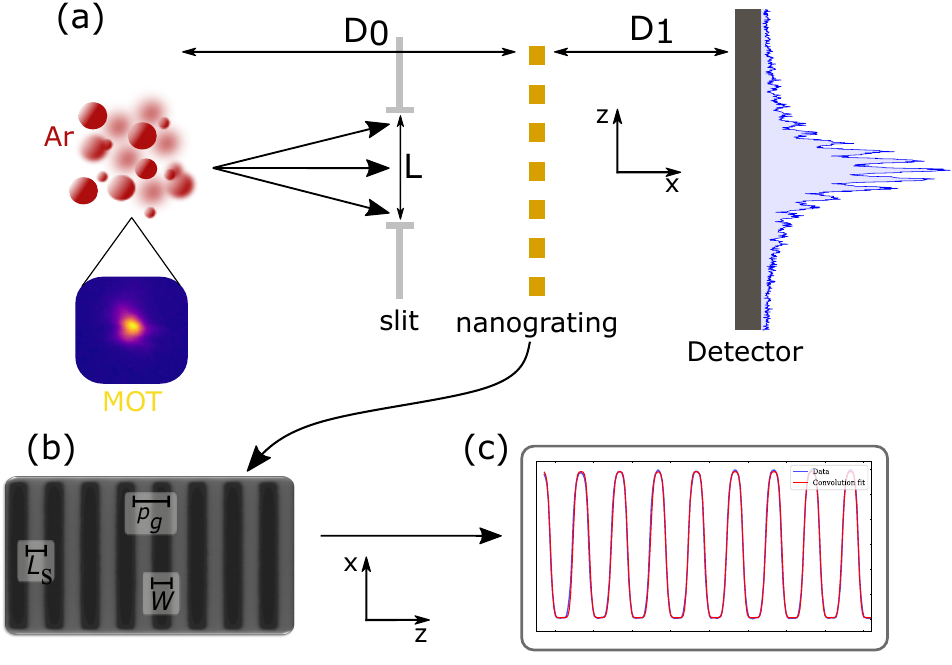}}
 \caption{\label{fig:exp_set_up}%
(a) Schematic of the experiment. An Argon MOT is driven toward the nanograting. A slit with an opening $L$ placed in front of the nanograting is utilized to select the angular dispersion of the atomic source. After interacting with the nanograting, the atoms are detected and the diffraction pattern is reconstructed. Given the nanoscopic size of the nanograting and the long interaction time, the diffraction image primarily reflects the influence of Casimir-Polder forces. In (b) is displayed a scanning electron microscope (SEM) image of the nanograting taken at 5 keV. $W$ denotes the slit width, $p_{g}$ the periodicity and $L_{s}$ the slit size. The slits are sufficiently large along the $x$ axis to disregard the diffraction pattern in that direction. (c) Cut of the SEM picture in the $x$ axis. The slit width is deduced by fitting the intensity profile of SEM images. The fit function is a convolution of a Gaussian function, representing the electron beam of the SEM, with a square function, representing an idealized nanograting geometry. The resulting slit width is determined to be $W = 97.0 \pm 3.6$\,nm. The uncertainty associated with the slit width comes from both the measured distribution of slit widths and the error associated with the SEM calibration (see appendix \ref{app:level1} for further details).}
\end{figure}

The experimental setup is depicted in Fig.~\ref{fig:exp_set_up}. The measurements are performed with noble Argon (Ar) atoms in the $\mathrm{^{3}P_{2}}$ metastable state. The use of these atoms ensures that the nanostructure is not subjected to chemical processes that might otherwise damage it, while still allowing for laser manipulation. The Ar atoms are initially laser-cooled in a magneto-optical trap (MOT), which serves as the atomic source for the interferometer. The MOT is composed of a quadrupole magnetic field with a gradient of 3.78\,G/cm (strong axis) and a light red detuned from the $\mathrm{^{3}P_{2} \rightarrow\,^{3}D_{3}}$ cycling transition by $\Delta=-2.5\,\Gamma$, with $\Gamma = 5.8$\,MHz. The intensity of the MOT light per beam is $I = 0.15\,I_{sat}$, with $I_{sat} = 0.14$\,mW/cm$^2$. This configuration creates an atomic source characterized by a Gaussian profile, with a width of $\sigma= 80\,\mu$m and a temperature of $T = 350\,\mu$K.

\begin{figure}[t]
{\includegraphics[width=\linewidth]{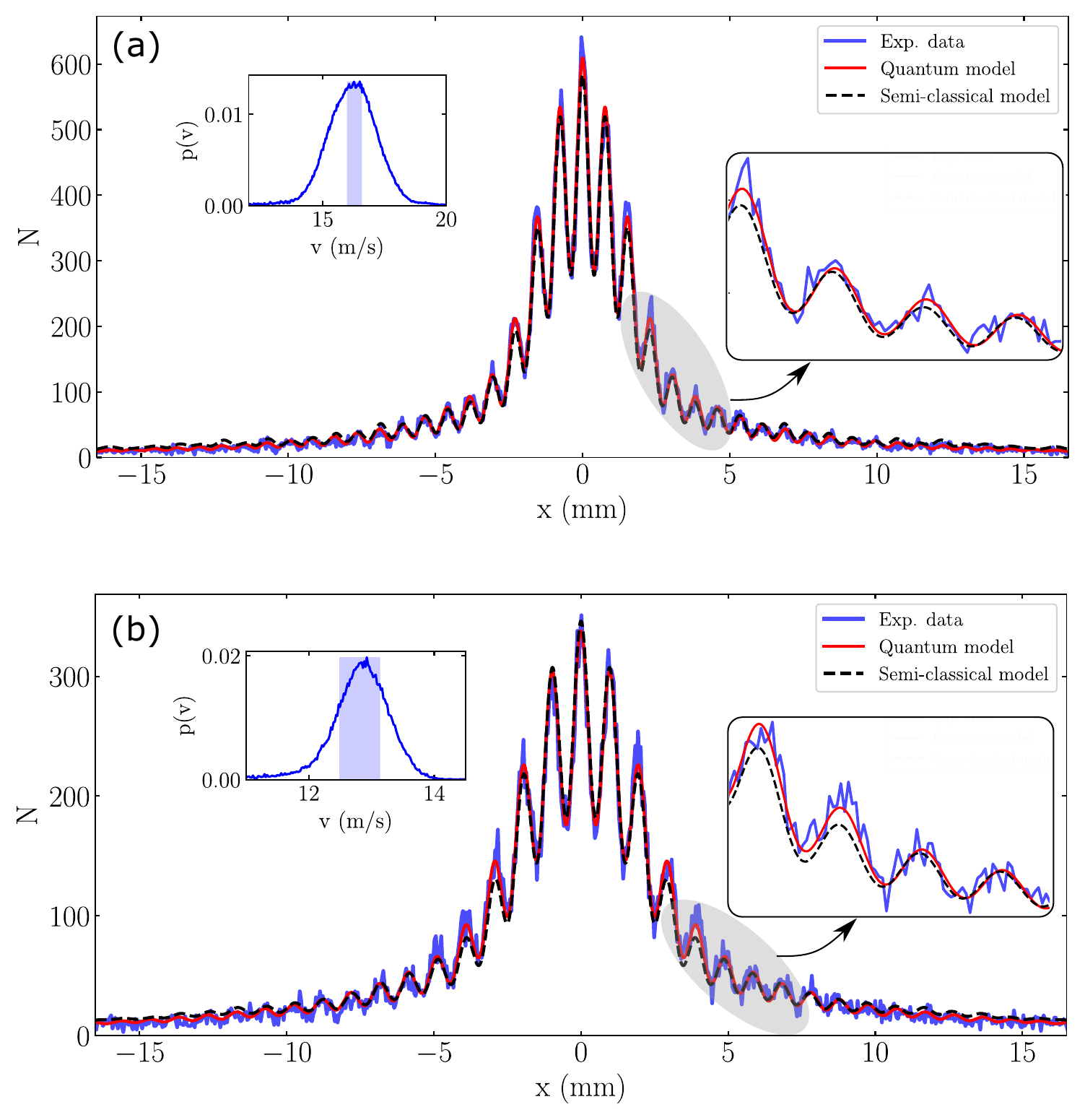}}
 \caption{\label{fig:diffraction_pattern}%
(a) In blue is plotted the diffraction pattern measured with velocities ranging from $v=15.99$\,m/s to $v=16.60$\,m/s. The red line represents the diffraction pattern calculated using the quantum model, while the black dashed line represents the diffraction pattern calculated using the semi-classical approach. A close-up of a particular region of the diffraction pattern is presented to illustrate the discrepancies between the two models. The measured velocity distribution is illustrated in the onset, with a vertical blue filled line indicating the selected velocity distribution. (b) is identical to (a) for the second set of data, which encompasses velocities ranging from $v=12.49$\,m/s to $v=13.14$\,m/s.}
\end{figure}

Thereafter, the MOT is switched off and the atoms are optically pushed toward the nanograting. The velocity of the atoms is determined through the use of a resonant light chopper technique, which enables the attainment of velocity measurement accuracies that fall below the percentage level. For the two diffraction spectra shown in Fig.~\ref{fig:diffraction_pattern}, we obtained mean velocities and respective uncertainties of $v= 16.26 \pm 0.05$\,m/s and $v = 12.81 \pm 0.02$\,m/s. Subsequently, after a propagation distance of $\mathrm{D_{0}}=56$\,cm, the atoms interact with a silicon nitride ($\mathrm{Si_3N_4}$) transmission grating, featuring a slit width of $W=97.0\pm 3.6$\,nm, a depth of $L_G=99.0$\,nm, and a periodicity of $p_g=200.0$\,nm, as depicted in Fig~\ref{fig:exp_set_up}. When the atoms pass through the nanograting, due to the scale involved they strongly interact with the slits via the C-P interaction. A mechanical slit positioned a few centimeters before the nanograting is used to control the angular distribution of the source.

The detection is then performed in the far-field regime at $\mathrm{D_{1}}=253.0 \pm 0.2$\,mm from the nanograting, using micro-channel plates in front of a delay line detector. This imaging setup allows for precise time-position detection, providing direct access to the longitudinal velocity distribution. The total duty cycle time is 100\,ms, resulting in a detected atomic flux of approximately 1.5 atoms per second. Due to the long interaction time between the atoms and the nanograting, the diffraction pattern is completely dominated by the C-P force. 

\section{\label{sec:level3}Theoretical model}

The C-P potential arises from the interaction energy between a fluctuating dipole and a nearby macroscopic body. This potential can be derived using perturbation theory applied to the atom-field Hamiltonian \cite{Buhmann2004}. Specifically, for half-space, the result can be written as \cite{Scheel2008}
\begin{align} \label{eqn:C.P_QED}
V_{C-P}(z) &  =  \frac{\hbar \mu_{0}}{8 \pi^{2}} \int_{0}^{+\infty} \mathrm{d}\omega\, \omega^{2} \alpha(i \omega) \int_{0}^{+\infty} \mathrm{d}k_{\parallel}\,\frac{k_{\parallel}}{\sqrt{k_{\parallel}^{2} + \frac{\omega^{2}}{c^{2}}}}  \nonumber \\
& \times \left( \mathrm{r^{TE}} - \left( 1+2 \frac{k_{\parallel}^{2}c^{2}}{\omega^{2}} \right) \mathrm{r^{TM}} \right) e^{-2z \sqrt{k_{\parallel}^{2} + \frac{\omega^{2}}{c^{2}}}}
\end{align}
where $\alpha(i \omega)$ is the Ar atom polarizability integrated over the imaginary frequencies, $\mathrm{r^{TE}}$ (respectively $\mathrm{r^{TM}}$) the transverse electric (respectively magnetic) reflection coefficients, and $z$ the atom-surface distance (see Appendix \ref{app:level2} for more details). This expression can be rewritten as follows   
\begin{equation} \label{eqn:C.P}
V_{C-P}(z) = \frac{C_{3}\,f(z)}{z^{3}}
\end{equation}
where the coefficient $C_{3}$ quantifies the strength of the interaction while the interpolation function $f(z)$ accounts for the onset of retardation effects caused by the finite field propagation time between the atom and surface ($f(z) \rightarrow 1$ when $z \rightarrow 0$). These effects become significant for distances greater than $\lambda / (2 \pi)$, where $\lambda$ is the wavelength of the optical transitions of the atom involved in the C-P potential calculation. In our study, the main relevant transitions occur at wavelengths $\lambda$=763, 811, and 912 nm, leading to retardation effects contributing at approximately 20\,\% to the total potential.

Far-field interferometry trough material nanograting is particularly well suited for exploring the C-P interaction \cite{Garcion2021,Perreault2005,Lepoutre2009}. In order to quantitatively analyze the impact of the C-P interaction on the diffraction pattern, we have recently developed a theoretical model based on a numerically efficient solution of the time-dependent Schr\"odinger equation \cite{Garcion2024}, going beyond the standard semi-classical approach. In short, we utilize the second-order split-operator technique inside the slits. The total potential considered is $V_{TOT}(z) = V_{C-P}(z) + V_{LJ}(z)$, where $V_{C-P}(z)$ is the C-P potential from Eq.\,(\ref{eqn:C.P}), and $V_{LJ}(z)= C_{rep}/z^9$ represents the repulsive Lennard-Jones interaction, resulting from Pauli repulsion between the electrons of atoms and those of the surface. The coefficient $C_{rep}$ describes the strength of this potential and is determined by fixing the position of the minimum potential at a distance $r_{min}$ from the slits walls. Additionally, a mask function is incorporated into the simulations to account for losses that occur during internal state transfer when Ar atoms come into contact with the surface \cite{Vassen2012}. Finally, the stationary phase approximation is employed to propagate the wavefunction from the slit output to the detector. Further details regarding this theoretical model can be found in Ref. \cite{Garcion2024}.  

\section{\label{sec:level4}Statistical analysis}

It is essential to exercise caution when comparing the theoretical model with the measurements, taking into account the potential for statistical and systematic errors. Statistical errors are the result of statistical uncertainties inherent to the data set, which may arise from stochastic fluctuations, finite statistical sampling, and the resolution of measurements. Systematic errors, on the other hand, are introduced by repeatable processes inherent to the system. Quantifying these errors in the measurements is crucial for utilizing the C-P effect as a test for exploring new short-distance forces \cite{Mostepanenko2016}. Here, we focus our analysis on the $C_{3}$ coefficient. Estimating this parameter is the primary objective of the statistical analysis, referred to as the point estimation. Denoting $\sigma_{stat}$ as the statistical error and $\sigma_{sys}$ as the systematic error associated to the $C_3$ coefficient, the total error is given by $\sigma = [\sigma_{stats}^{2} + \sigma_{sys}^{2}]^\frac{1}{2}$.

\subsection{$\chi^{2}$ analysis}

Our discussion begins with the treatment of statistical errors, an essential step in the data analysis process. These errors place limits on the accuracy of the information that can be extracted from the diffraction pattern. Two measured diffraction spectra are shown in Fig~\ref{fig:diffraction_pattern}. The longitudinal velocity distribution is determined for each data set, with values chosen within the ranges $v \in [15.99,16.60]$\,m/s for the first set of data and $v \in [12.49,13.14]$\,m/s for the second set of data. The first set of data comprises $N_{1}=54.4 \times 10^{3}$ atoms, while the second set comprises $N_{2}=44.1 \times 10^{3}$ atoms. This distribution is incorporated into the theoretical models. The data binning was chosen to ensure that there are 10 data points for each diffraction peak, thus preventing any issues with over-sampling or sub-sampling. Minor detector aberrations have been corrected to reconstruct the diffraction pattern using a 3600-point grid pattern.\\

The limited precision and finite data points call for the application of a statistical $\chi^{2}$-fit analysis. This analysis is necessary to determine whether our data support or refute the theoretical model. It is thus a prerequisite step that must be completed before any data interpretation can be made. Given the nature of the detected diffraction patterns and the small number of atoms per bin in the tails, a $\chi^{2}$ for a multinomial distribution \cite{baker_clarification_1984} is utilized to identify the best fit parameter which minimizes
\begin{equation}
\chi^{2}(C_3) = 2 \sum_{i=1}^{N_{bin}} n_{exp}(x_{i}) \ \mathrm{ln} \left( \frac{n_{exp}(x_{i})}{n_{theo}(x_{i}, C_3)} \right),
\end{equation}
where $n_{exp}(x_{i})$ [resp. $n_{theo}(x_{i}, C_3)$] is the experimental [resp. theoretical] atom number at the position $x_{i}$, and $N_{bin}$ is the data sampling.

In the following, we present reduced $\chi_{red}^{2}$ values defined as $\chi_{red}^{2}(C_{3}) = \chi^{2}(C_3)/N_{bin}$. The best-fit results for both the quantum model we have developed \cite{Garcion2024} and the standard semi-classical method are displayed in Fig.~\ref{fig:diffraction_pattern}. The semi-classical approach is described with classical waves and the atomic wavefunction is estimated by means of action integrals along classical trajectories \cite{Fiedler2022}. In the first data set ($v \in [15.99,16.6]$\,m/s, $N_{1}=54.4 \times 10^{3}$ atoms and $N_{bin}=743$), the quantum model yields a minimum $\chi_{red}^{2}=1.07$ with a point estimation of $C_{3}=3.50$ $\mathrm{meV.nm^{3}}$, while the semi-classical model results in a minimum $\chi_{red}^{2}=2.10$ with $C_{3}=2.87$ $\mathrm{meV.nm^{3}}$. Similarly, in the second data set ($v \in [12.49,13.14]$\,m/s, $N_{2}=44.1 \times 10^{3}$ atoms and $N_{bin}=743$), the quantum model gives a minimum $\chi_{red}^{2}=1.06$ for $C_{3}=3.75$ $\mathrm{meV.nm^{3}}$, while the semi-classical approach results in a minimum $\chi_{red}^{2}=1.80$ with $C_{3}=3.51$ $\mathrm{meV.nm^{3}}$ respectively. In both sets of experimental data, we find that the quantum model produces a lower $\chi_{red}^{2}$ compared to a semi-classical approach, indicating that the data are in favor of the quantum model.\\

\begin{figure}
{\includegraphics[width=\linewidth]{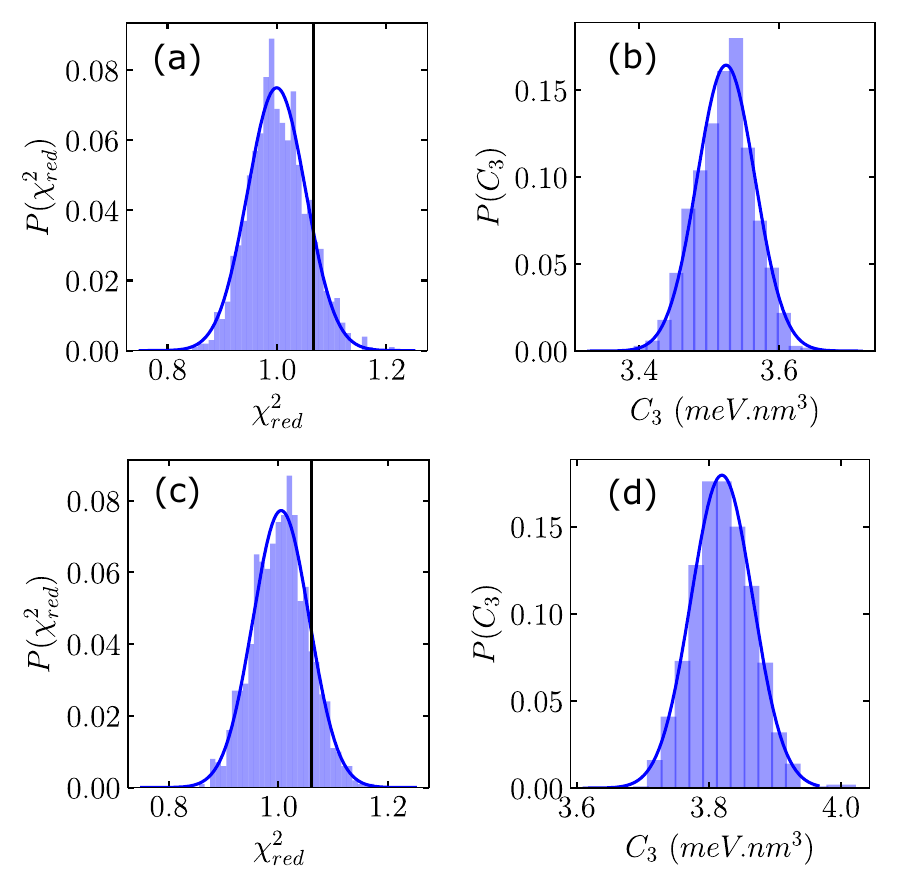}}
 \caption{\label{fig:chi2_MC}%
(a) $\chi_{red}$ distribution derived from the Monte-Carlo simulations for the data with an average velocity of 16.26\,m/s using the quantum model. The vertical black line represents the value obtained from experimental data. This value is within the distribution's range, indicating that the quantum model is consistent with our data. (b) $C_{3}$ distribution extracted from the Monte-Carlo simulations. The width of such distribution reveals the relative statistical error, resulting in $\sigma_{stats}/C_{3} = 1.2 \%$. (c) and (d) are the same as (a) and (b) for the data with mean velocity 12.81\,m/s.}
\end{figure}

To evaluate the models, we perform a goodness-of-fit test. To achieve this, we reconstruct the $\chi_{red}^{2}$ distribution using Monte Carlo simulations based on von Neumann's acceptance-rejection algorithm. Typically, we generate simulated data from the model with the best-fit parameters, maintaining the same total number of atoms as in the experimental data. We then identify the $C_{3}$ values that minimize $\chi_{red}^{2}$ for the simulated data. By repeating this process multiple times, we obtain the typical distribution of $\chi_{red}^{2}$ expected if the experiment conforms to the theoretical model. The result is shown in Fig.~\ref{fig:chi2_MC}. The $\chi_{red}^{2}$ distribution can be fitted with a Gaussian having a mean of 1 and a standard deviation of $2 \sigma = 0.1$. Based on this distribution, to determine the acceptance or rejection of our models, i.e. to determine the boundaries of $\chi_{red}^{2}$ where the models either accept or reject the observed data, we use the $P$-value \cite{P_value_remark}. This quantity indicates the probability of observing $\chi_{red}^{2}$ values similar to those obtained from our experimental data in $P\times 100\,\%$ of the cases. We have established a threshold of $P=0.02$ for model acceptance, indicating that the model will be accepted if $\chi_{red}^{2} \leq 1.1$ (corresponding to $2 \sigma$ of the distribution). We have verified that the goodness-of-fit test results remain consistent when increasing or reducing the binning.

\begin{figure}[t]
{\includegraphics[width=\linewidth]{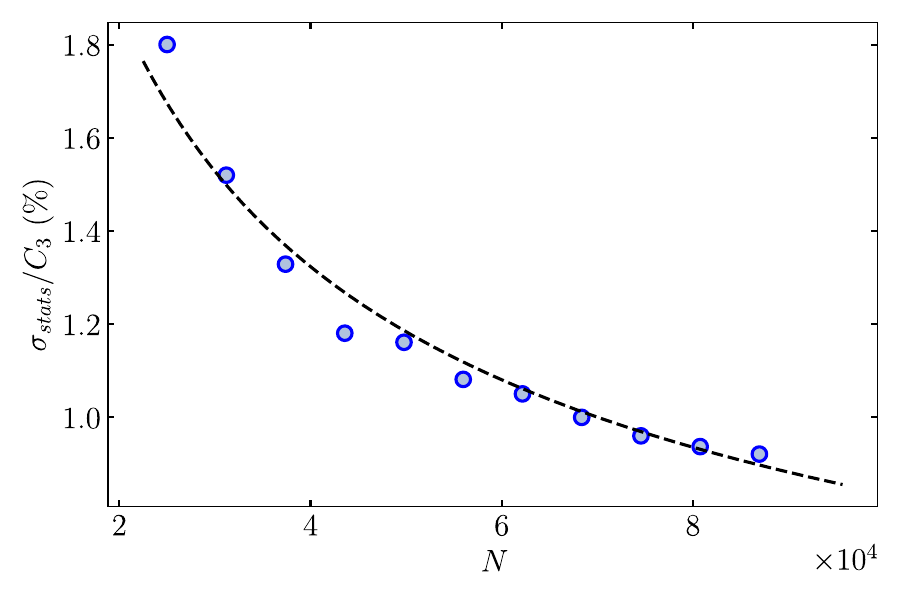}}
\caption{\label{fig:Quantum_noise_limitation}%
Relative statistical error $\sigma_{stats}/C_{3}$ as a function of the total atom number $N$. The black dashed line is a fit to the data proportional to $1/\sqrt{N}$, showing that our measurements are limited by the quantum shot noise.}
\end{figure}

The Monte-Carlo procedure also allows us to extract the $C_{3}$ distribution. The $C_{3}$ distribution for the quantum model is plotted in Fig.~\ref{fig:chi2_MC}. This distribution follows a Gaussian distribution where the width directly reflects the statistical error $\sigma_{stats}$. For both datasets, we find $\sigma_{stats}/C_{3} = 1.2 \%$. This number directly depends on the statistics and thus on the total atom number $N$. Accumulating data points from the second set of data with velocities comprising between $v=12.49$\,m/s and $v=13.14$\,m/s, we study the dependency of $\sigma_{stats}$ with $N$. The results are presented in Fig.~\ref{fig:Quantum_noise_limitation}. We observe that $\sigma_{stats}/C_{3}$ decreases as $1/ \sqrt{N}$, as anticipated for independent measurements. This suggests that our measurement is primarily limited by quantum shot noise rather than experimental flaws. This also implies that we would need four times more data to reduce the relative random error by a factor of 2. The current atom number used in our data ($N \approx 50 \times 10^{3}$) is a trade-off between statistical significance and measurement duration (almost 9 hours for each measurement).  

Furthermore, a bootstrapping procedure was employed on both datasets to corroborate the statistical dispersion inferred by the Monte-Carlo method, and to validate the confidence intervals found. We obtain identical results. More specifically, the $\chi_{red}^{2}$ distribution exhibits the same Gaussian shape and width, resulting in the same statistical errors. In addition, we observe that the decrease in $\sigma_{stats}/C_{3}$ is also proportional to $1/\sqrt{N}$. These findings increase our confidence in the estimated statistical errors of our measurements and in our methodology.

\subsection{Atomic source}

\begin{figure}[t]
{\includegraphics[width=\linewidth]{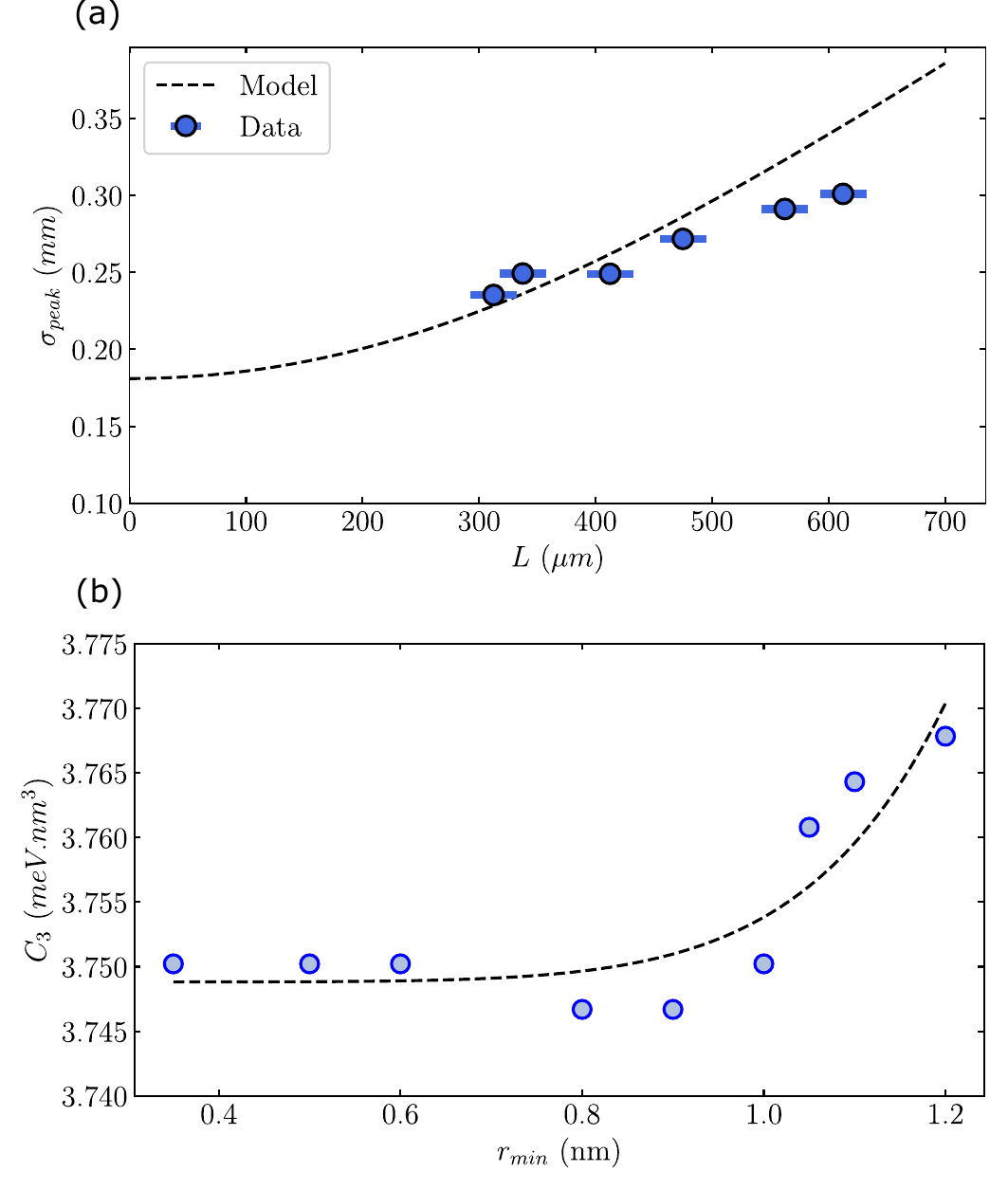}}
 \caption{\label{fig:Source_size_and_LJ}%
(a) Diffraction peak width $\sigma_{peak}$ as a function of the mechanical slit opening $L$. The solid points represent experimental data points, which were directly extracted from diffraction spectra with Gaussian fits. For each data point, the longitudinal velocity distribution was chosen such as it has a negligible impact on $\sigma_{peak}$. The dashed line corresponds to the model described in the text and in appendix \ref{app:level3}. The finite value of $\sigma_{peak}$ when $L \rightarrow 0$ corresponds to the finite size source effect. The model and the data are in good agreement. (b) Influence of the Lennard-Jones parameter $r_{min}$ on the estimation of the C-P parameter $C_{3}$. The black-dashed line is a fit that serves as a guide for the eye.}
\end{figure}

The typical width $\sigma_{peak}$ of the diffraction peaks is an essential parameter to estimate when comparing data to the model. In this context, we study and characterize this experimental parameter. The finite width of the diffraction peak originates from two primary factors: the spatial extent of the source (related to the van Cittert theorem), and the longitudinal $(k_{z})$ and transversal $(k_{x})$ momentum distribution of the source (related to van Cittert-Zernike theorem). The longitudinal momentum distribution is measured with the imagery set up and is incorporated in the model. In standard atomic interferometry experiment, the spatial extension of the source and its transversal $(k_{x})$ momentum distribution are usually controlled by two consecutive slits positioned in front of the nanogratings \cite{Bruhl2002, Hornberger2012}. This configuration ensures high spatial coherence, which is necessary to distinguish small diffraction angles when working with supersonic jets\,\cite{Grisenti2000}. In our case, due to the low atomic flux, only a single mechanical slit is positioned a few centimeters before the nanograting (see Fig.~\ref{fig:exp_set_up}). Therefore, we investigate experimentally in Fig.~\ref{fig:Source_size_and_LJ}(a) how the peak width $\sigma_{peak}$ is affected by the slit opening $L$ in our unconventional situation. As expected, we observe that larger values of $L$ correspond to larger peak widths, an effect that can be attributed to the increase in selected transverse momenta.

To be more quantitative, we now compare this result with a simple analytical model. This model assumes that the atomic source follows a Gaussian distribution. The atomic cloud reaches its final velocity after exchanging a number of photons $N_{\gamma}$, equal to the ratio of the atomic velocity, $v$, to the recoil velocity for a wavelength of 811\,nm, $v_{rec}$. This leads to $N_{\gamma}$ cycles of absorption and spontaneous emission. These optical cycles result in a random walk distribution, which can be described for each atom by a Gaussian transverse velocity distribution with a width of $\sigma_{s} = \sqrt{N_{\gamma}/3}\,v_{rec}$. Subsequently, this distribution is truncated due to the presence of the slits. In this context, a Gaussian distribution with a width of $\sigma_{peak}$ provides an accurate model for the diffraction peaks. Further details regarding the model can be found in appendix \ref{app:level3}. In Fig.~\ref{fig:Source_size_and_LJ}(a), we compare this model with the data and we observe that the model is in good agreement with the experimental points. Additionally, an independent measurement of the width of the transverse momentum distribution was conducted for both datasets, and the results were found to be in agreement with the widths extracted from the model at a $5 \%$ level. This slight difference can be attributed to the small non-parallelism of the mechanical slits. In conclusion, this parameter is well under control in our experimental configuration, and thus does not impact the C-P information extracted from the experimental data, provided that $\sigma_{peak}$ remains significantly smaller than the interfringe.

\subsection{Lennard-Jones potential}

At close proximity to the surface, the atoms become sensitive to the repulsive force caused by the interaction between the electrons of the atoms and those of the surface. This force is described using a Lennard-Jones potential, expressed as follows: $V_{LJ}(z)=C_{rep}/z^{9}$. The value of $C_{rep}$ represents the strength of this potential and is determined by setting the location of the minimum potential at a distance $r_{min}$ from the slits walls. This potential plays a role in adsorption processes on surfaces \cite{Bruch2007}, but remains challenging to characterize and probe experimentally \cite{Debiossac2014}. It has been proposed to tailor this potential to capture atoms near the surface, with the ultimate goal of developing nanoscale atom-surface metamaterials \cite{Chang2014,Whittaker2014}. It is therefore important to search for a signature of this potential. To the best of our knowledge, no investigation of this potential has yet been conducted using our method. The quantum model that supports our data naturally incorporates this potential \cite{Garcion2024}, allowing us to explore its impact on the diffraction pattern. By focusing on the second data set with a mean velocity centered at $v=12.81$\,m/s, we vary $r_{min}$ and examine the best estimation for $C_{3}$. The results are plotted in Fig.~\ref{fig:Source_size_and_LJ}(b). It is notable that there is a slight sensitivity of the diffraction pattern to the Lennard-Jones potential, amounting to approximately 0.5\,\% in the $C_{3}$ coefficient. This small signature can be attributed to the fact that short-range physics inside the slits is translated into signals in the tails of the diffraction pattern, where only a few atoms are present. Nevertheless, this sensitivity remains minor in comparison to the additional systematic effects present in our measurements, as will be discussed in the following section.

\section{\label{sec:level5} Correction factors and systematic error analysis}

The quantum model that corroborates the data can be employed to investigate the systematic effects and errors that are present in our experiment. Systematic effects result in a shift in the estimated parameter $C_{3}$, while experimental uncertainties lead to systematic errors $\sigma_{C3}$ on the $C_{3}$ coefficient.

\subsection{Slit width $W$}

\begin{figure}[t]
{\includegraphics[width=\linewidth]{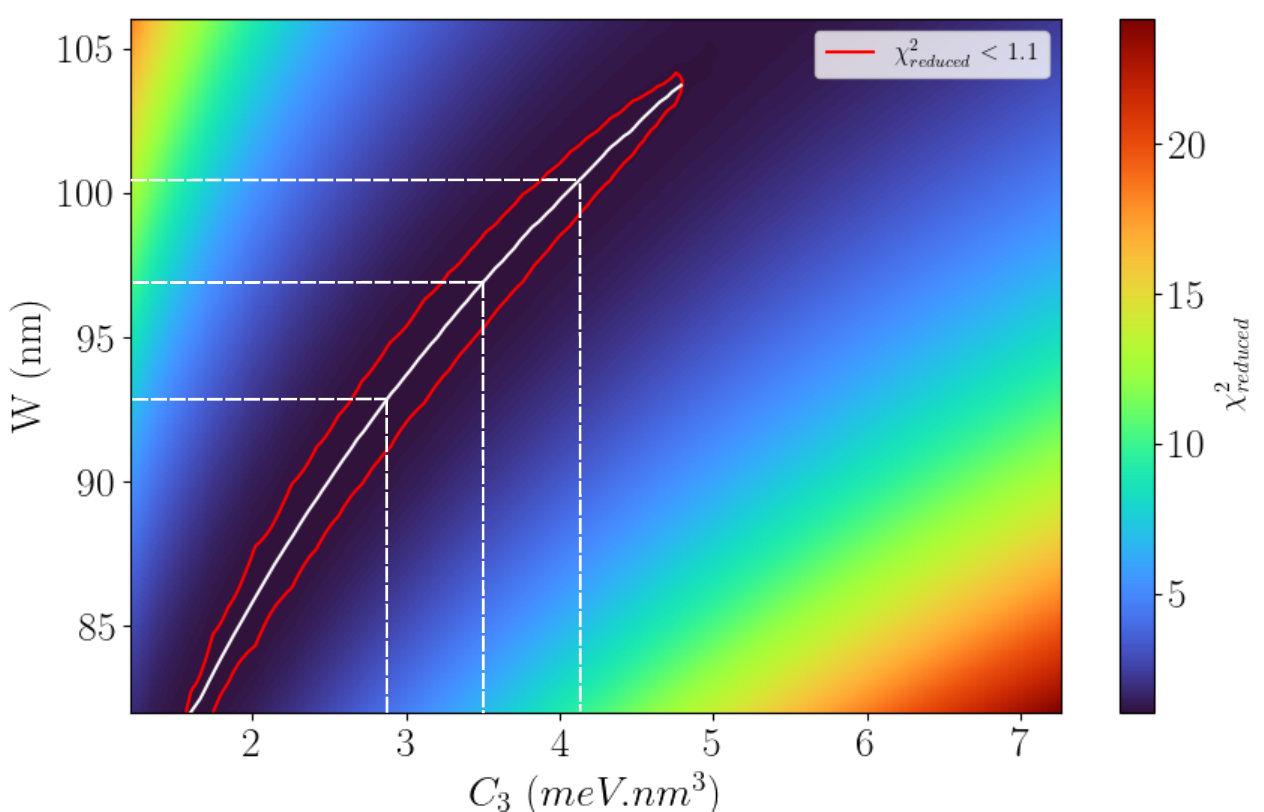}}
\caption{\label{fig:geometry}%
2D graph representing $\chi_{red}^{2}$ as a function of the coefficient $C_{3}$ and the width $W$. The atomic velocity is set at an average value of $v = 16.26$\,m/s. The red contour plot indicates the area where the quantum model is accepted, corresponding to accepted couples $\{C_{3}, W \}$. We observe that a range of values cannot be excluded from the model, entailing the strong correlation between $C_3$ and $W$. The horizontal white dashed lines represent the uncertainty in the slit width measurement, which has been determined to be $W= 97.0 \pm 3.6$\,nm. This leads to a systematic error of $\sigma_{geo}/C_{3} = 14\,\%$, as indicated by the horizontal white dashed lines.}
\end{figure}

The geometric parameters of the nanograting, including depth $L_G$ and slit width $W$, could exert a significant influence on the determination of the $C_{3}$ coefficient. The depth of the nanograting has been accurately measured through ellipsometry, and is found to be $L_{G}=99.0$\,nm, with error bars below 1\,nm. The following section will discuss the influence of $W$ on the $C_3$ coefficient. The width $W$ is determined using conventional scanning electron microscope (SEM) imagery. However, this method, like transmission electron microscopy imagery, is affected by charge effects and is hindered by calibration issues during each measurement. Other techniques, such as scanning tunneling microscopy or atomic force microscopy, depend on a convolution of the sample and the unknown tip shape. Consequently, one of the main sources of uncertainty in the point estimation parameter $C_{3}$ is the lack of precise knowledge of the nanograting slit width $W$. For the diffraction pattern, a slight increase in the coefficient $C_{3}$ is analogous to a slight increase in the width $W$. This relationship has been explored through interferometry experiments utilizing a supersonic beam  \cite{Perreault2005,Lonij2009} and/or employing a semi-classical approach \cite{Garcion2021}. However, the precise nature of this effect remains only partially understood. In this study, we investigate the dependency of the phenomenon in question by varying the width $W$ in our quantum model and comparing the theoretical results to the first set of data, applying the $\chi_{red}^{2}$ procedure described previously. The results are displayed in Fig.~\ref{fig:geometry}. We observe that our model is accepted (i.e $\chi_{red}^{2} \leq 1.1$) for a range of coupled values $\{C_{3}, W\}$, highlighting the strong correlation between these two parameters. Our SEM fitting analysis and procedure yield an uncertainty of $\Delta W = 3.6$\,nm, which leads to a systematic error of $\sigma_{geo}/C_{3} = 14\,\%$ (see Fig.~\ref{fig:geometry}, and appendix \ref{app:level1} for further details). This value represents the main limitation in our error budget, as detailed later in Table\,\ref{tab:error_budget}. 

\subsection{Finite slit dimensions}

\begin{figure}[ht]
{\includegraphics[width=\linewidth]{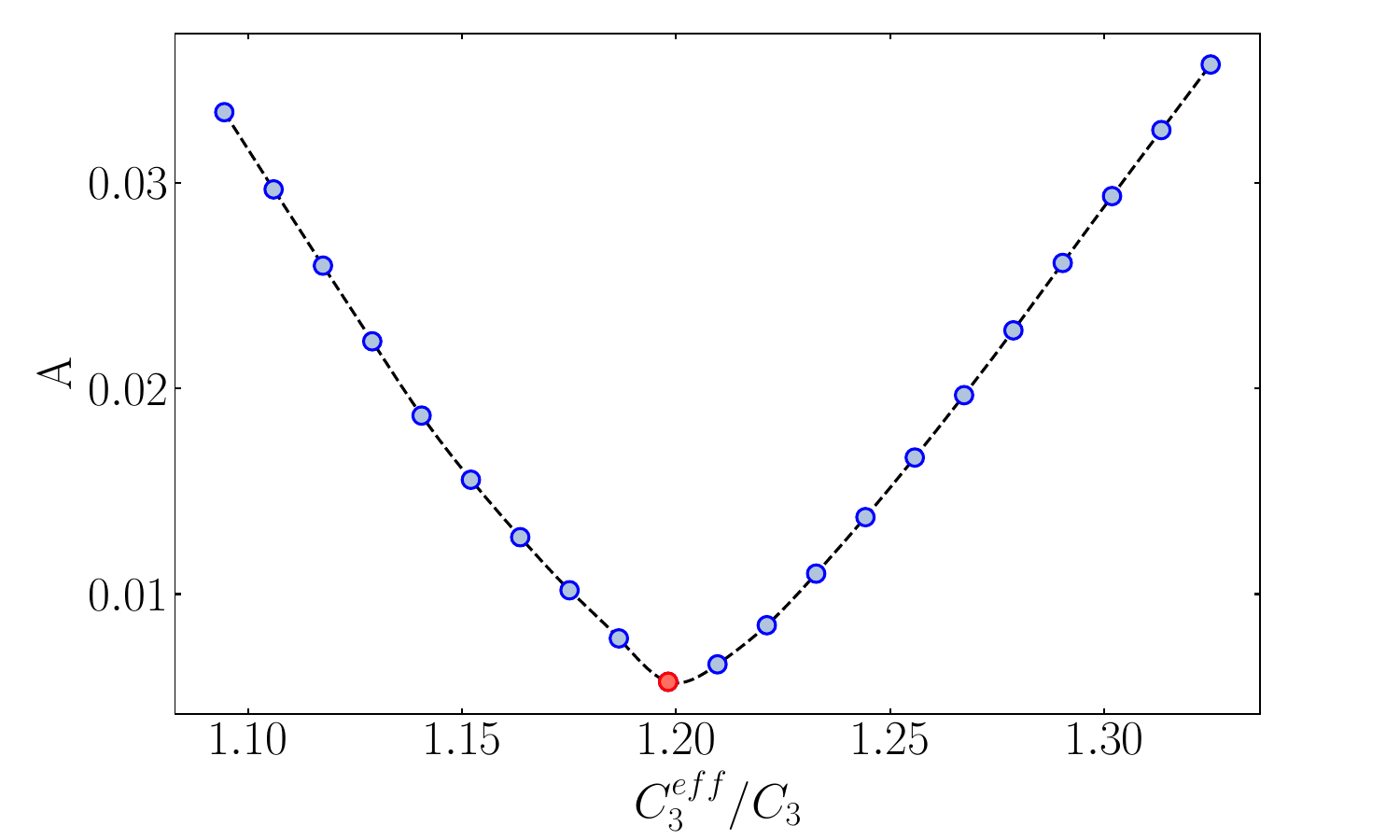}}
\caption{\label{fig:scanc3}%
Dimensionless indicator $A$ as a function of the ratio $C_3^{eff}/C_{3}$ for a finite grating depth $L_{G} = 99.0$\,nm. The average atomic velocity is set at $v = 16.26$\,m/s. The red point denotes the minimum of $A$, which occurs at a value of 0.006 for $C_3^{eff}/C_{3} = 1.20$. The black-dashed line is a fit that serves as a guide for the eye.}
\end{figure}

The nanograting has a finite depth of $L_{G}= 99.0$ nm, whereas the theoretical model discussed in section \ref{sec:level3} assumes an infinite planar surface. Consequently, a smaller grating surface interacts with the Ar atoms. This feature has two consequences. Firstly, it scales the previously inferred $C_{3}$ coefficient. Secondly, this finite size introduces, through the uncertainty on the width $W$, a systematic error of finite dimension noted $\sigma_{f.d}$. To quantify both effects, we compare the results of the simulations between a grating with semi-infinite walls and a grating with finite dimensions. For the purposes of this analysis, we focus on the envelope of the diffraction pattern, which contains the C-P information. To compute the resulting C-P potential with finite dimensions, we utilize the pair-wise approximation, which can be expressed analytically as $V(x,z) = C_{3}^{eff} g(x,z) / z^3$ inside the slits (see appendix\,\ref{app:level5} for further details). This two-dimensional expression of the C-P. potential is introduced into our quantum model\;\cite{Garcion2024} via the correspondence $x = v t$, which transforms this time-independent two-dimensional potential into a time-dependent one-dimensional potential. In this context, $C_{3}^{eff}$ represents the strength of the atom-surface interaction, while the function $g(x,y)$ accounts for the finite dimension of the grating. This expression does not account for retardation effects. Consequently, in the case of semi-infinite walls, we only compute Eq.\,(\ref{eqn:C.P}) in the near-field limit, defined as $2 z \omega /c \leq 1$. In this scenario, the potential is represented by the expression $V(z)=C_{3}/z^{3}$ (derived from Eq.\,(\ref{eqn:C.P}) with $f(z)=1$), where the parameter $C_{3}$ corresponds to the values inferred previously by the statistical $\chi^2_{red}$ analysis. Both approaches yield identical results when $L_{G}$ is sufficiently large. To infer the shift in the coefficient $C_{3}$ resulting from the finite depth $L_{G}$, we proceed as follows: firstly, the envelope of the diffraction pattern for a semi-infinite surface $|\psi_{s.i}(x, C_{3})|^2$ is simulated in the near field limit. Subsequently, the envelope of the diffraction pattern in the finite dimension pair-wise model, $|\psi_{f.d}(x, C_3^{eff})|^2$, is simulated while varying $C^{eff}_{3}$. A comparison between the two models is then made using the following dimensionless indicator
\begin{equation}\label{eq:indicator}
A = \frac{\displaystyle\int \left|\big|\psi_{s.i}(x, C_{3})\big|^2-\big|\psi_{f.d}(x, C_3^{eff})\big|^2\right|dx}{\displaystyle\int \big|\psi_{s.i}(x, C_{3})\big|^2\,dx}\,.
\end{equation}   
Therefore, $A$ quantifies the relative difference between the two models. The lower $A$ is, the closer the two diffraction patterns are. The result is shown in Fig~\ref{fig:scanc3}. We observe that the finite dimension causes a correction factor of 1.20 in the original $C_{3}$ coefficient for atoms with mean velocities $v=16.26$\,m/s. The same coefficient of 1.20 is found for atoms with mean velocities $v=12.81$\,m/s. Hence, we have $C_{3} = 4.2$\,meV.nm$^3$ for the first dataset and $C_{3} = 4.5$\,meV.nm$^3$ for the second dataset. The systematic error arises from the uncertainty in the slit width $W=97.0 \pm 3.6$ nm. By implementing the $W$ extremes values in the simulation, we find that the systematic error due to the finite grating depth $\sigma_{f.d}/C_{3}$ is far below 1 $\%$ and can be thus considered negligible.

\subsection{Opening angles}
  
\begin{figure}[t]
{\includegraphics[width=\linewidth]{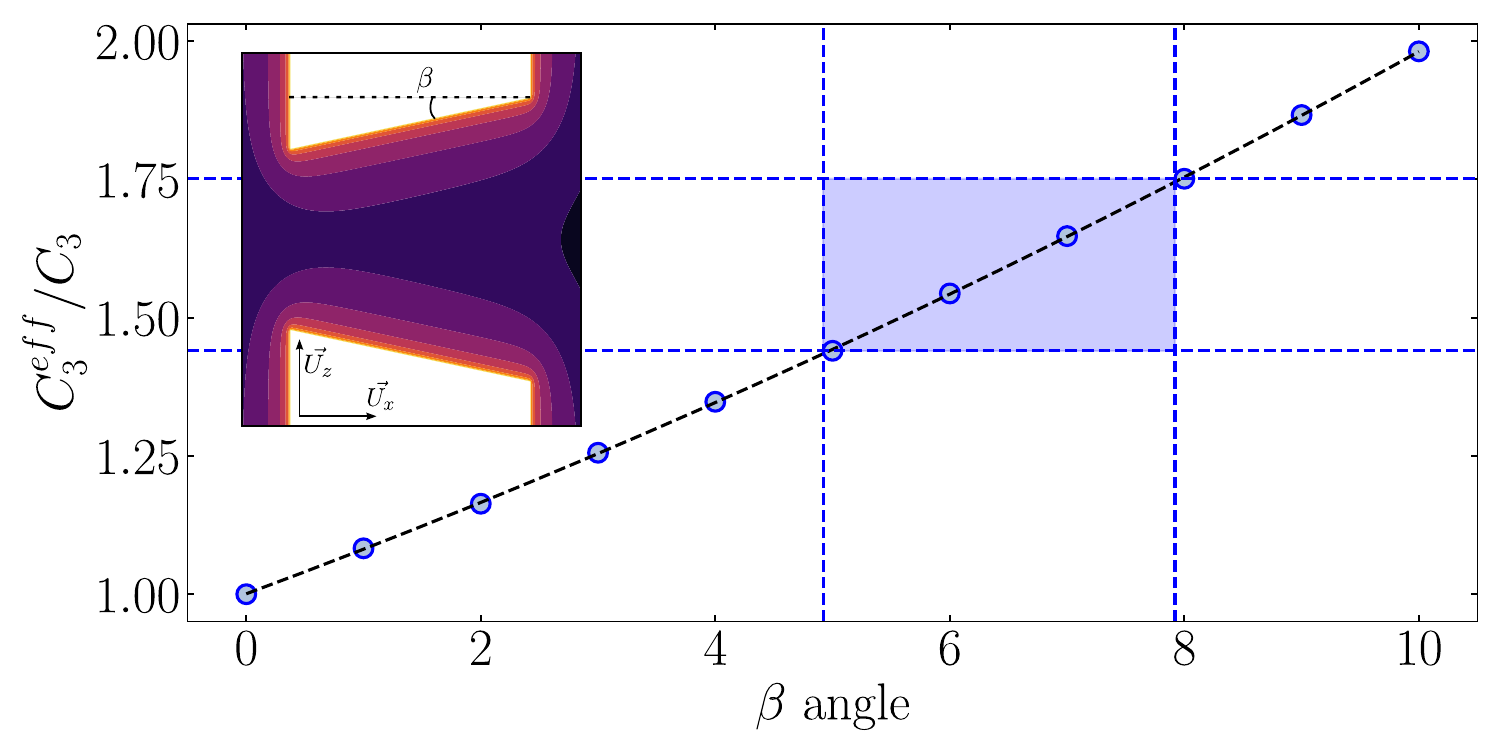}}
 \caption{\label{fig:scanbeta}%
Ratio $C_3^{eff} / C_{3}$ as a function of the opening angle $\beta$, for an atomic average velocity fixed at $v = 16.26$\,m/s. The black-dashed line is a fit that serves as a guide for the eye. The purple rectangular area indicates the uncertainty interval of the angle $\beta$, thus representing the systematic error $\sigma_{\beta}$.  The insert presents a 2D plot of the pair-wise C-P potential with opening angle $\beta$.
}%
\end{figure}

A detailed analysis of the nanograting geometric parameters via SEM reveals the existence of an opening angle, designated as $\beta$, with a mean value of $(6.4 \pm 1.5)$\,degrees, as illustrated in Fig.~\ref{fig:scanbeta}. The uncertainty associated with the opening angle is directly related to the error bars on the measured front and back slit widths. This introduces an additional correction in the $C_{3}$ coefficient and a systematic error $\sigma_{\beta}$ due to the uncertainty on the angle $\beta$. The aim of this section is to evaluate both effects. In order to compute them, we initially simulate the envelope of the diffraction pattern $|\psi_{f.d}(x, C_{3})|^2$ for parallels walls and finite dimensions using the pair-wise approximation. With the same approximation, the expected envelope of the diffraction pattern $|\psi_{\beta}(x, C_3^{eff})|^2$ is then computed when the nanograting possesses an opening angle of $\beta$ and a C-P coefficient $C_3^{eff}$ (see appendix \ref{app:level4} for more details). The comparison between the probability densities $|\psi_{f.d}(x, C_{3})|^2$ and $|\psi_{\beta}(x, C_3^{eff})|^2$ is then realized using a new indicator $B$ analogous to the one defined in Eq.\,(\ref{eq:indicator})
\begin{equation}\label{eq:indicatorB}
B = \frac{\displaystyle\int \left|\big|\psi_{f.d}(x, C_{3})\big|^2-\big|\psi_{\beta}(x, C_3^{eff})\big|^2\right|dx}{\displaystyle\int \big|\psi_{f.d}(x, C_{3})\big|^2\,dx}\,.
\end{equation}
For a fixed value of $\beta$, we first vary the $C_3^{eff}$ coefficient in order to find the one that minimizes $B$. The $C_3^{eff}$ coefficient found quantifies the extent to which the $C_3$ parameter must be corrected due to the opening angle. We then repeat the same procedure for different values of $\beta$. The outcome is presented in Fig.~\ref{fig:scanbeta}: A strong correlation is observed between the $C_3^{eff}$ coefficient and the opening angle, $\beta$. For a $\beta$ value of 6.4\,degrees, the opening angle leads to a correction factor of 1.58 in the $C_{3}$ coefficient for both data sets since we could verify that the correction factor is not influenced by the choice of average velocity. The corrected $C_{3}$ point estimation is thus $C_{3}= 6.63$\,meV.nm$^3$ for the first dataset with mean velocity $v=16.26$\,m/s and $C_{3}= 7.11$\,meV.nm$^3$ for the second dataset with mean velocity $v=12.81$\,m/s. Finally, the uncertainty of the opening angle $\beta$ results in a systematic error $\sigma_{\beta} / C_{3} = 9.85\,\%$ (see Fig.~\ref{fig:scanbeta}). This substantial value emphasizes the strong relationship between our measurements and the knowledge of the geometric parameters of the nanograting.

\subsection {Conclusion on systematic effects}

We have also investigated other systematic effects, such as the finite size of the slits. When comparing the C-P potential $V_{C-P}$ for a semi-infinite surface to a layer with thickness $L_{s}=99.0$\,nm (see Fig~\ref{fig:exp_set_up}), we find no discernible difference. As a result, we conclude that the systematic effects $\sigma_{L}$ arising from the finite slit size are negligible (see appendix \ref{app:level2}). Additionally, the nanograting is estimated to have a roughness of approximately 1-2\,nm. Since our measurement occurs in the far-field regime and such effect are dominant at short atom-surface distances, we estimate this effect to be negligible compared to others \cite{Garcion2024}. Finally, the relation between the $C_{3}$ coefficient and the width $W$ being non-linear, an uncertainty $\Delta W$ leads also to a shift in the $C_{3}$ coefficient. We consider such a shift as a systematic error $\sigma_{N.L}$. We finds $\sigma_{N.L}$ = 0.7 $\% C_{3}$. In conclusion, the total error is given by $\sigma = [\sigma_{stats}^{2} + \sigma_{sys}^{2}]^\frac{1}{2} = 17.2\,\%$ $C_{3} $, where $\sigma_{sys}^{2} \approx \sigma_{geo}^{2} + \sigma_{\beta}^{2}$. All the errors are outlined in the error budget Table \ref{tab:error_budget}.

\begin{table}[h!]
\centering
\begin{tabularx}{0.45\textwidth} { 
  >{\centering\arraybackslash}X 
  >{\centering\arraybackslash}X}
\hline
\hline
Quantity & Experimental uncertainty in $C_{3}$ ($\%$)   \\
\hline
\hline
$\sigma_{stats}$  & 1.2   \\
\hline
$\sigma_{geo}$  & 14   \\
\hline
$\sigma_{f.d}$  & $<0.5$   \\
\hline
$\sigma_{\beta}$  & 9.9   \\
\hline
$\sigma_{L_{s}}$  & $<0.1$   \\
\hline
$\sigma_{N.L}$  & 0.7   \\
\hline
\hline
\end{tabularx}
\caption{\label{tab:error_budget}%
Sources of error for the measurement of the point estimation parameter $C_{3}$.}
\end{table}

\section{\label{sec:level6}General conclusion}

This study has successfully validated a quantum numerical model of atomic diffraction by a material nanograting. This validation enables us to investigate both statistical and systematic effects on our measurements. By employing statistical tools, including the reduced $\chi^{2}_{red}$ and Monte-Carlo methods, we demonstrate that the statistical error on the C-P potential strength parameter $C_{3}$ is 1\,\%. This error is limited only by shot noise. Furthermore, we have investigated the impact of the atomic source and the sensitivity of the short-range Lennard-Jones potential. Additionally, we have examined how the slit width ($W$), the opening angle ($\beta$), and the finite depth size ($L_{G}$) influence the C-P interaction, revealing a significant impact on the $C_{3}$ coefficient. The lack of precise knowledge about these geometric parameters leads to significant systematic errors in determining the value of $C_{3}$ (on the order of 17.2\,\%), emphasizing the strong connection between the C-P potential and the nanograting geometry. Our findings indicate that $C_{3} = 6.87 \pm 1.18$\,meV.nm$^3$ (average of two data sets presented in this study). This value is consistent with the expected theoretical value $C_{3}=5.04$\,meV.nm$^3$, which corresponds to the Lifshitz formula, i.e. when $f(z) \rightarrow 1$ \cite{Lifshitz1956,Mavroyannis1963}. The precision of this value strongly depends on the susceptibility of the nanograting material (see appendix \ref{app:level2}). This can be a limitation in the infrared domain, where susceptibility measurements are scarce. Furthermore, the value also depends on stoichiometry and deposit conditions, which can alter the refractive index and introduce uncertainties in the predicted theoretical value. For instance, a 20\,\% alteration in the dielectric response, denoted as $\epsilon(i \omega)$, gives rise to a 10\,\% shift in the $C_{3}$ coefficient. Such modifications are still within the realm of possibility in our experiment. Therefore, the accurate estimation of uncertainty remains a very challenging task. Finally, the experimental value obtained in this study is also consistent with other experimental values reported in previous studies: $C_{3} = 5.00$\,meV.nm$^3$ in \cite{Garcion2021}, $C_{3} = 7.42$\,meV.nm$^3$ in \cite{Morley2021}, and $C_{3} = 7.37$\,meV.nm$^3$ in \cite{Karam2005}. However, these studies did not consider or acknowledge the impact of nanograting geometries on the C-P potential. Furthermore, the model used to interpret their data was of a semi-classical nature and not purely quantum.

The lack of precise knowledge regarding the nanograting geometry represents the primary limitation of the measurement accuracy in this work. Therefore, the development of new methods to enhance the accuracy of the measurements is a promising avenue for future research. Modifying the angle of incidence between the atomic beam and the nanograting could assist in constraining and decoupling the nanograting geometry from the C-P potential \cite{Lonij2009}. Coupled with tomographic methods \cite{Stielow2021}, this approach would be sensitive to the precise form of the C-P potential, potentially enabling tests of gravity at short distances and providing new constraints on a hypothetical non-Newtonian fifth force \cite{Antoniadis2011}. 

\begin{acknowledgments}

LPL is UMR 7538 of CNRS and Sorbonne Paris Nord University. The authors gratefully acknowledge the support and computing resources from the the MAGI HPC center at Université Sorbonne Paris Nord. The authors are grateful to  the French technological network RENATECH and the platform IEMN (UMR 8520 of CNRS) for nanogratings manufacturing and characterizations. We also would like to thank Maria Konstantakopoulou from LSPM (UPR 3407 of CNRS) for nanograting images. We wish to acknowledge Benoît Darquié for fruitful discussions.  This work has been partially supported by structure fédérative de recherche NAP MOSAIC of the University Sorbonne Paris Nord. N.G. acknowledges funding from the Deutsche Forschungsgemeinschaft (German Research Foundation) under Germany’s Excellence Strategy (EXC-2123 QuantumFrontiers Grants No. 390837967) and through CRC 1227 (DQ-mat) within Projects No. A05, and the German Space Agency at the German Aerospace Center (Deutsche Raumfahrtagentur im Deutschen Zentrum f\"ur Luft- und Raumfahrt, DLR)  with funds provided by the German Federal Ministry of Economic Affairs and Climate Action due to an enactment of the German Bundestag under Grants Nos. 50WM2250A and 50WM2250E (QUANTUS+) and No. 50WM2253A (AI-Quadrat).

\end{acknowledgments}

%%%%%%%%%%%%%%%%%%%%%%%%%%%%%%%%%%%%%%%%%%%%%%%%%%%%%%%%%%%%
%\newpage
\revappendix

\section{\label{app:level1}  Fabrication process and image analysis of the Si$_3$N$_4$ nanograting}

The manufacturing process of nanograting involves a number of essential steps. Initially, E-beam resist is directly applied by spin-coating on a pre-defined Silicon Nitride 100 nm thick membrane to allow direct patterning of the nano-grating. Then, Reactive Ion Etching is employed to create the grating into the membrane through the mask designed by the E-beam lithography process. Finally, cleaning procedures such as resist dissolution and RF O2 plasma ashing are carried out to remove the resist and residual material. \\

\begin{figure}[t]
    \centering
    \includegraphics[width=\linewidth]{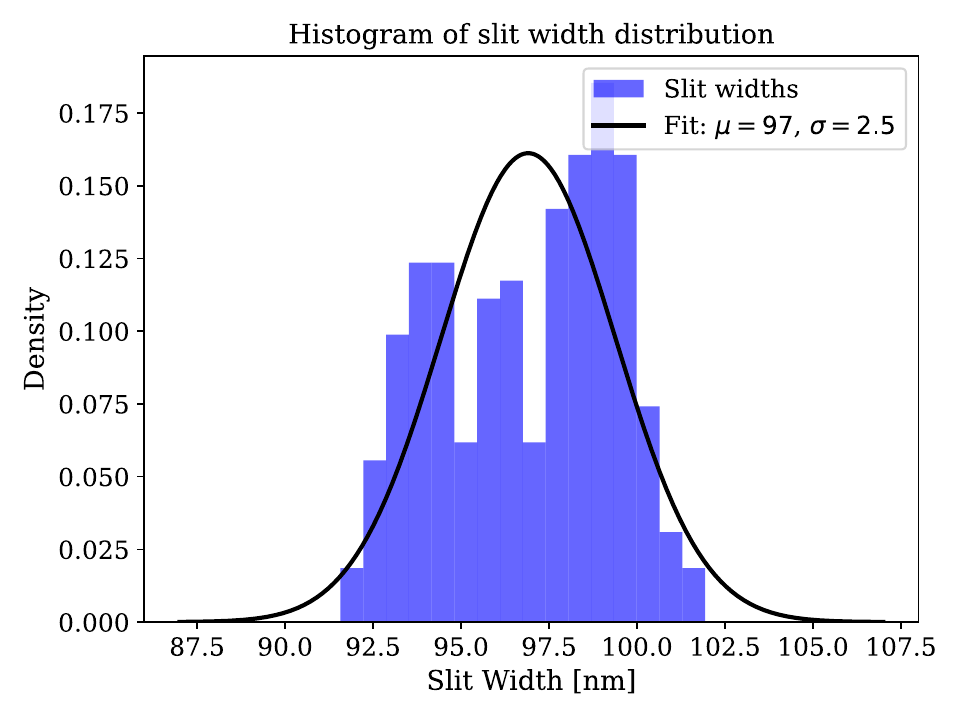}
    \caption{In blue, measured width slit distribution of the nanograting. The black line represents the Gaussian fit.
    }
    \label{width_distribution}
\end{figure}

The nanogratings produced were characterized by standard scanning electron microscope (SEM) imagery. This technique is generally constrained by charge effects, magnetic lens aberrations and electron diffraction. To mitigate the charge effects while achieving a significant penetration depth of the electron beam, images are taken with an electron beam energy of 5\,kV. Aberration effects and limited resolution due to electron diffraction are assumed to follow a Gaussian profile. The SEM images are therefore fitted by the convolution of two functions. The first function is a Gaussian function, which represents the spread point function of the SEM. The second function is a square function that defines the ideal nanograting geometry. By analyzing multiple randomized samples of SEM images of the nanograting, we are able to deduce the slit width distribution (see Fig.~\ref{width_distribution}). The mean slit width is found to be $W = 97.0$\,nm with a standard deviation of $\sigma_{W} = 2.5$\,nm. Furthermore, the error associated with SEM calibration is estimated to be $\sigma_{cal} = 2.6$\,nm. Consequently, the total error on the slit size is $\sigma = [\sigma_{W}^2+\sigma_{cal}^2]^\frac{1}{2} = 3.6$\,nm. This leads to the final value of the width, $W = (97.0 \pm 3.6)$\,nm.

%%%%%%%%%%%%%%%%%%%%%%%%%%%%%%%%%%%%%%%%%%%%%%%%%%%%%%%%%%%%
\section{\label{app:level2} Casimir-Polder potential}

\begin{figure}[t]
{\includegraphics[width=0.9\linewidth]{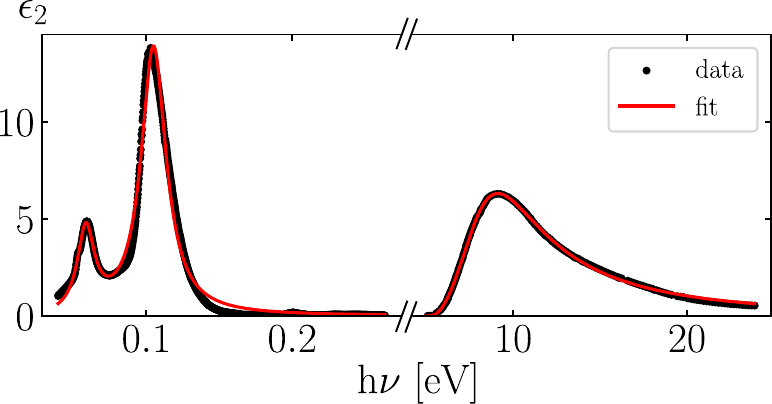}}
 \caption{\label{fig:IR+UV}%
Plot of the imaginary part $\epsilon_{2}$ of the dielectric function of $\mathrm{Si_{3}N_{4}}$ in black and plot of the Tauc-Lorentz fit in red.}%
\end{figure}

In the formulation of the C-P. potential given in Eq.\,(\ref{eqn:C.P_QED}), the transverse electric field coefficient $\mathrm{r^{TE}}$ and the magnetic reflection coefficient $\mathrm{r^{TM}}$ play crucial roles. The $\mathrm{r^{TE}}$ coefficient between air and $\mathrm{Si_{3}N_{4}}$ writes
\begin{equation}
\mathrm{r^{TE}}(i \omega, k_{\parallel}) = \frac{\beta_{0}-\beta_{SiN}}{\beta_{0}+\beta_{SiN}}\,,
\end{equation}
while $\mathrm{r^{TE}}$s can be expressed as 
\begin{equation}
\mathrm{r^{TM}}(i \omega, k_{\parallel}) = \frac{\epsilon(i \omega) \beta_{0}-\beta_{SiN}}{\epsilon(i \omega) \beta_{0}+\beta_{SiN}}\,,
\end{equation}
where $\epsilon(i \omega)$ is the relative dielectric function of $\mathrm{Si_{3}N_{4}}$ on the imaginary axis. The terms $\beta_{0}$ and $\beta_{SiN}$ are defined as
\begin{eqnarray}
\beta_{0}   & = & i \sqrt{\frac{\omega^{2}}{c^2} + k_{\parallel}^{2}}\,,\\
\beta_{SiN} & = & i \sqrt{\frac{\epsilon(i \omega)  \omega^{2}}{c^2} + k_{\parallel}^{2}}\,.
\end{eqnarray}
The dielectric function $\epsilon(i \omega)$ can be deduced using the Kramers-Kronig relation
\begin{equation}\label{KramersKronigIm}
\epsilon(i \omega)= 1+\frac{2}{\pi}\int_{0}^{\infty} \frac{t  \epsilon_{2}(t)}{t^2+\omega^2} dt\,,
\end{equation}
where $\epsilon_{2}=\Im(\epsilon)$ is the imaginary part of the dielectric function $\epsilon$. The quantity $\epsilon_2$ is inferred using the Tauc-Lorentz model in order to fit the optical properties in the ultra-violet (wavelengths from 50\,nm to 250\,nm) \cite{Philipp1973} and in the infrared (between 290\,nm and 30\,$\mu$m) \cite{Luke2015}. It thus leads to $\epsilon_2 = \epsilon_{2,UV} + \epsilon_{2,IR}$ with 
\begin{equation}
\label{eqn:TaucLoUV}
\epsilon_{2,UV}(\omega)= \Theta(\omega-E_{g,UV})\,\frac{A E_0 \Gamma (\omega-E_{g,UV})^2}{[(\omega^2-E_0^2)^2+\Gamma^2\omega^2]\omega}\,,
\end{equation}
and
\begin{equation}
\label{eqn:TaucLoIR}
\begin{split}
\epsilon_{2,IR}(\omega) = & \Theta(\omega-E_{g,IR})
\left[\frac{A_1 E_{1} \Gamma_1 (\omega-E_{g,IR})^2}{[(\omega^2-E_{1}^2)^2+\Gamma_1^2\omega^2]\omega} \right. \\[0pt]
& \qquad\quad\quad + \left. \frac{A_2 E_{2} \Gamma_2 (\omega-E_{g,IR})^2}{[(\omega^2-E_{2}^2)^2+\Gamma_2^2\omega^2]\omega} \right]\,,
\end{split}
\end{equation}
where $\Theta$ is the Heaviside step function, and $E_{g,IR}$, $A$, $E_0$, $E_{g,UV}$, $A_1$, $E_1$, $\Gamma_1$, $A_2$, $E_2$, $\Gamma_2$ are the fit parameters. The numerical values of these fit parameters are given in table \ref{tab:TableFitParam} and the imaginary part of the dielectric function $\epsilon_2$ is plotted in Fig.\ref{fig:IR+UV}.

\begin{table}[h!]
\centering
\begin{tabularx}{0.45\textwidth} { 
  >{\centering\arraybackslash}X 
  >{\centering\arraybackslash}X}
\hline
\hline
 Fit parameter  & numerical value (in eV)\\
\hline
\hline
 $\hbar A$  & 251.5\\
\hline
$E_0$ &  8.1\\
\hline
$\hbar \Gamma$ & 5.8\\
\hline
 $E_{g,UV}$ &  5.2\\
 \hline
$\hbar A_1$ &  0.3368\\
\hline
$E_1$ &  0.1054\\
\hline
$\hbar \Gamma_1$ & 0.0208\\
\hline
$\hbar A_2$ &  0.0811 \\
\hline
$E_2$ &  0.0589 \\
\hline
$\hbar \Gamma_2$ &  0.0141 \\
\hline
$E_{g,IR}$ &  0.0081 \\ 
\hline
\end{tabularx}
\caption{Numerical values of the Tauc-Lorentz fit parameters for the imaginary part $\epsilon_{2}$ of the dielectric function of $\mathrm{Si_{3}N_{4}}$.}
\label{tab:TableFitParam}
\end{table}

\begin{figure}[t]
{\includegraphics[width=\linewidth]{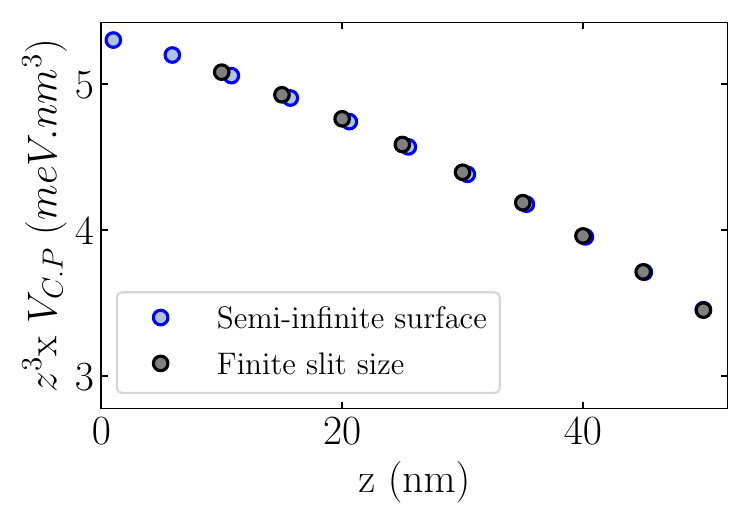}}
 \caption{\label{fig:CP_potential_plot}%
$z^{3} \times V_{C-P}$ as a function of the atom-surface distance $z$ for the case of a semi-infinite surface (blue point) and for the finite slit size $L_{s}=100$ nm (black point). The difference between the two potentials is negligible. }%
\end{figure}

The C-P potential $V_{C-P}(z)$ calculated from the imaginary dielectric function $\epsilon_{2}$ depicted in Fig.~\ref{fig:IR+UV} is shown in Fig.~\ref{fig:CP_potential_plot}. To account for the finite size $L_{s}$ of the nanograting slit, the coefficients $\mathrm{r^{TM,TE}}$ must be modified to include the effects of multiple reflections at the interfaces \cite{Fiedler2019} 
\begin{equation}
\mathrm{r_{L_s}^{TM,TE}} = \frac{r^{TM,TE}}{1-r^{TM,TE} e^{-2i\beta_{SiN} L }}
\end{equation}
The potential $V_{C-P}$ considering the finite size of the slits is displayed in Fig.~\ref{fig:CP_potential_plot}. The difference between this potential and the potential $V_{C-P}$ calculated assuming a semi-infinite surface is extremely small, and thus does not play a significant role in our configuration.

%%%%%%%%%%%%%%%%%%%%%%%%%%%%%%%%%%%%%%%%%%%%%%%%%%%%%%%%%%%%
\section{\label{app:level3}Source model}

The presence of a single mechanical slit positioned before the nanograting (see Fig.~\ref{fig:exp_set_up}) allows us to select the angular distribution $\theta$ of the atomic source. Large values of slit opening $L$ correspond to an increase in the selected transverse momenta $\theta$, resulting in less visibility in the diffraction spectrum. To understand the dependency between these two quantities, we develop a simple analytical model. Initially, the source is assumed to have a Gaussian distribution with a size $\sigma_{0}$. Each atom of the distribution is then pushed by a quasi-resonant light to reach a final longitudinal velocity $v$. During this process, atoms exchange many photons, leading to a Gaussian velocity distribution for each atom of the source with a size $\sigma_{t} = \sqrt{v v_{rec}/3}\,t$, where $v_{rec}$ is the recoil velocity of the atom at the wavelength 811\,nm, and $t$ is the time the atoms propagate from the source to the detector. The distance from the source to the detector is called $D$. The position $x$ of the atom on the detector and the angular distribution $\theta$ are connected by $\theta=x/D$. This distribution, consisting of atoms evolving with a Gaussian distribution at $\sigma_{t}$ average over the initial source position (Gaussian distribution at $\sigma_{0}$), is then truncated at $D_{0}$ by the slits. In other words, the slits select specific angular angle $\theta$. The source distribution $S(\theta)$ on the detector thus writes
\begin{equation}
S(\theta)=\int_{\Theta - L/2D_{0}}^{\theta + L/2D_{0}} \exp{ \left(\frac{-D^{2}}{2\sigma^{2}_{t}} \left( \theta - \frac{D_{1} \alpha}{D} \right)^{2} -\frac{D_{1}^{2} \alpha^{2}}{2 \sigma_{0}^{2}} \right)} \, d\alpha 
\end{equation}

Computing this integral, we find 
\begin{equation}
\begin{aligned}
S(x)= & \exp{\left( - \frac{x^{2}}{2 \sigma^{2}_{t}} \left(  1-\frac{1}{1+\sigma^{2}_{t}/\sigma^{2}_{0}} \right)\right) } \times \\
&   \left( \mathrm{erf} \left( A \left( \frac{x}{D} + \frac{L}{2D_{1}} - B \frac{x}{D_{1}} \right) \right) \right.   \\
&    - \left. \mathrm{erf} \left( A \left( \frac{x}{D} - \frac{L}{2D_{1}} - B \frac{x}{D_{1}} \right) \right) \right) 
\end{aligned}
\end{equation}
where $\mathrm{erf}$ is the error function, $A^{2} = D_{1}^{2}/2 \left( 1/ \sigma^{2}_{t} + 1/ \sigma^{2}_{0} \right)$ and $B= 1/(1+\sigma^{2}_{t}/\sigma^{2}_{0})$. This distribution can be accurately described by a Gaussian distribution with a width of $\sigma_{peak}$ as long as the slit opening $L$ is not significantly larger than the initial source size, which is indeed the case in the experiment. In Fig.~\ref{fig:Source_size_and_LJ}, we compare this model with the experimental data. We observe that the model agrees with the experimental points. In conclusion, this parameter is effectively controlled in our experimental setup and does not affect the extraction of C-P information from the experimental data as long as $\sigma_{peak}$ is smaller than the interfringe distance.

%%%%%%%%%%%%%%%%%%%%%%%%%%%%%%%%%%%%%%%%%%%%%%%%%%%%%%%%%%%%
\section{\label{app:level4} Pair-wise potential for a rectangular slit}

\begin{figure}[t]
{\includegraphics[width=\linewidth]{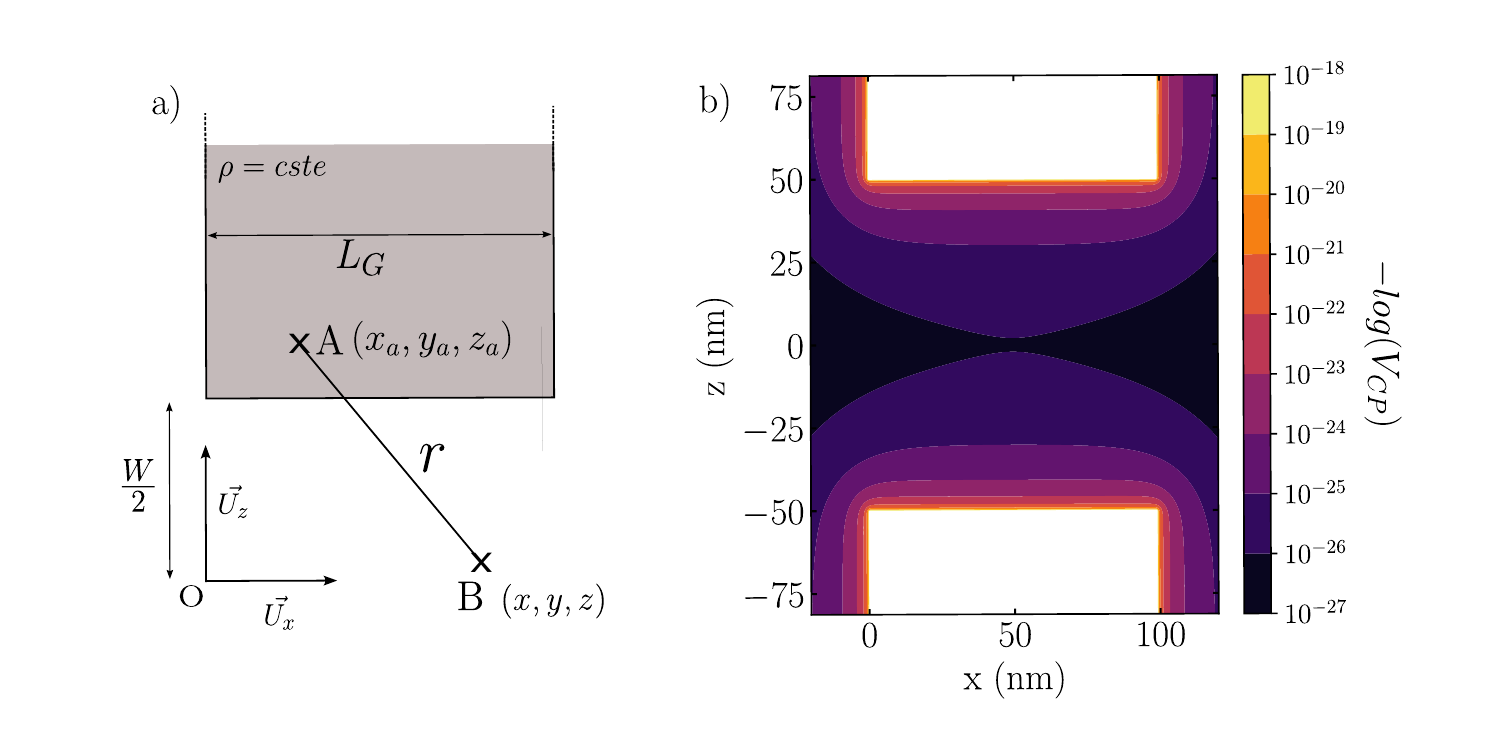}}
 \caption{\label{fig:pote_finite}%
(a) Coordinate system used in the calculation of the pair-wise calculation for a rectangular slit. Here, $\rho$ is the atomic density of the grating, $L_G$ the slit depth, $W$ slit width and $r$ the distance between an Argon atom and an atom inside the nano-grating. (b) Plot of the potential $V_{CP}(x,z)$ in log scale.}%
\end{figure}

To compute the C-P potential with finite slit dimensions, i.e with a depth $L_{G}$ and a width of $W$ (refer to Fig.~\ref{fig:pote_finite}), we use the pair-wise approximation, where the atom-surface interaction is expressed as a summation of Van der Waals potential between all the atoms within the surface and the atom. This approach thus neglects collective effects in the materials \cite{Bitbol2013}. The Van Der Walls interaction between two atoms is written as $V_{VDW} = - C_{6}/r^6$, where $r$ is the distance between the two atoms and $C_{6}$ the strength of the interaction. Assuming a constant atomic density $\rho$ inside the material and an invariance along the y-axis (see Fig.~\ref{fig:pote_finite}), the pair-wise potential is for the upper slit at $z=W/2$
\begin{equation}
V_{CP}(x,y,z)=\int_{W/2}^{+\infty} \int_{-\infty}^{+\infty} \int_{0}^{L_G} -\frac{\rho C_6}{r^6} \, dx_a \, dy_a \, dz_a
\label{eqn:firsttripleint}
\end{equation}
where $r=\sqrt{(x-x_a)^2+(y-y_a)^2+(z-z_a)^2}$. This integrals can be calculated analytically and we obtain:  
\begin{equation}
\begin{aligned}
&V_{CP}(x,y,z) = \frac{\rho C_6 \pi}{12}\bigg[ \frac{1}{x^3} + \frac{1}{(L_G-x)^3}  \\
& +\frac{x^4+x^2(z-W/2)^2/2+(z-W/2)^4}{x^3(z-W/2)^3\sqrt{x^2+(z-W/2)^2}} \\
&+ \frac{(L_G-x)^4+(L_G-x)^2(z-W/2)^2/2+(z-W/2)^4}{(L_G-x)^3(z-W/2)^3\sqrt{(L_G-x)^2+(z-W/2)^2}} \bigg] \,
\end{aligned}
\label{eqn:CP_finite}
\end{equation}

\begin{figure}[t]
{\includegraphics[width=\linewidth]{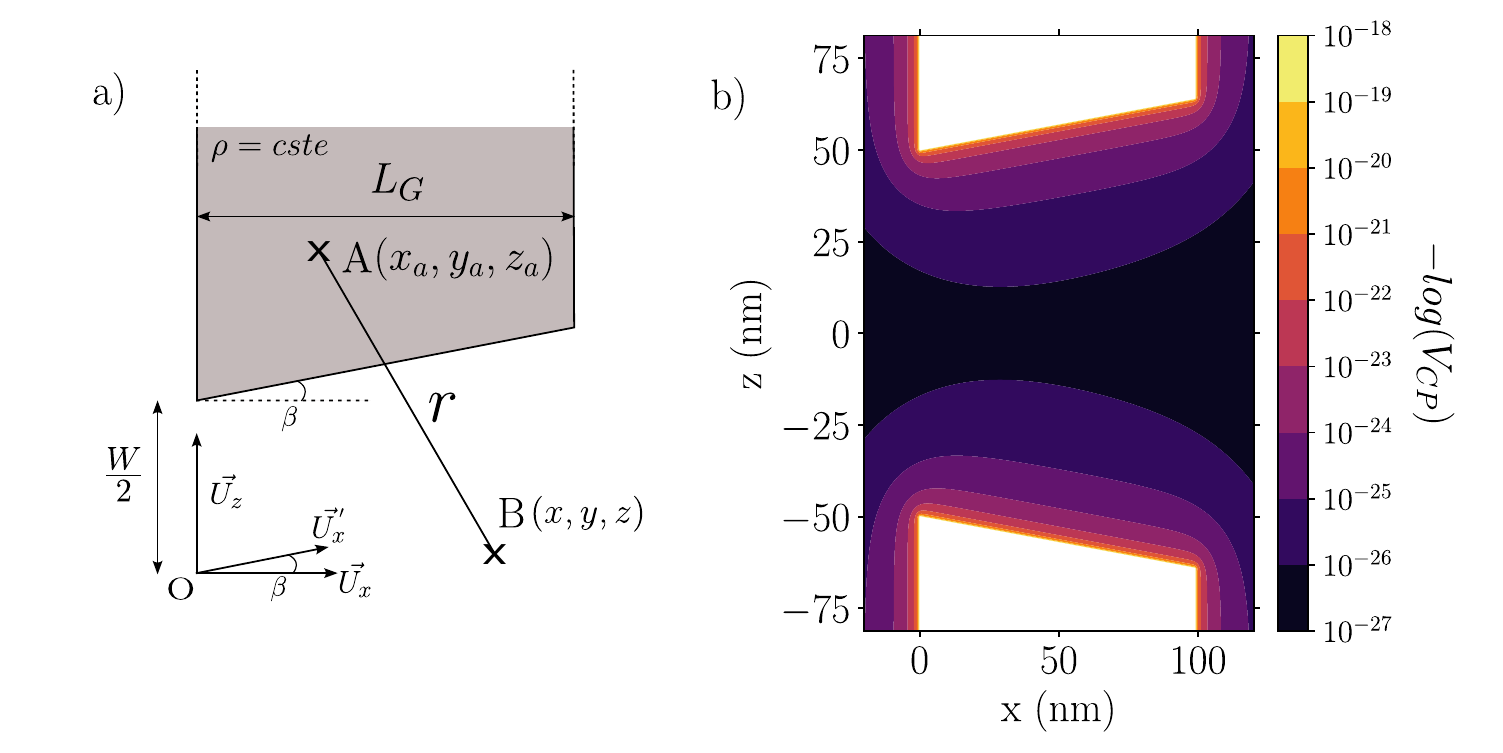}}
 \caption{\label{fig:pote_finite_trapez}%
(a) Coordinate system used to calculate the pair-wise potential of a trapezoidal-shaped slit. (b) Graph of the potential $V_{CP}(x,z)$ on a logarithmic scale.}%
\end{figure}

After translating the potential from $x \rightarrow x + L_G/2$, we note that, when $ L\rightarrow +\infty$, we recover the usual C-P potential without retardation effects $V_{CP}(z)=C_3/(z-W/2)^3$, where $C_3=\rho C_6\pi/6$. The pairwise potential for the lower slit at $z=-W/2$ can be easily calculated by replacing $z$ with $-z$ in the expression (\ref{eqn:CP_finite}). In fig~\ref{fig:pote_finite}, the total pair-wise potential is depicted.

%%%%%%%%%%%%%%%%%%%%%%%%%%%%%%%%%%%%%%%%%%%%%%%%%%%%%%%%
\section{\label{app:level5} Pair-wise potential for a trapezoidal slit}

A detailed analysis of the nanograting shows the presence of an opening angle $\beta$, leading to a trapezoidal slit shape (see Fig~\ref{fig:pote_finite_trapez}). In order to compute the C-P potential for this geometry, we continue to use the pair-wise approximation method that is described as follows for the upper slit at $z=W/2$ :
\begin{equation}
V_{CP}(x,y,z)=\int_{W/2 +x_a \tan(\beta)}^{+\infty} \int_{-\infty}^{+\infty} \int_{0}^{L_G} -\frac{\rho C_6}{r^6} \, dx_a \, dy_a \, dz_a
\end{equation}
To facilitate the calculation, we implement the following variable change
\[
\begin{cases}
    x' = x/\cos(\beta) \\
    z' = z - x\tan(\beta)
\end{cases}
\]
The C-P potential, which has now independent integration limit, is expressed as   
\begin{equation}
\begin{aligned}
V_{CP}(x',y',z')= &-\rho C_6\int_{W/2-z'}^{+\infty} \int_{-\infty}^{+\infty} \int_{-x'}^{L_G/\cos(\beta)-x'} \\
&\frac{1} {{(x_a'^2+y_a'^2+z_a'^2+2x_a'z_a'\sin(\beta))}^3} \, dx_a' \, dy_a' \, dz_a'
\end{aligned}
\end{equation}
We can calculate this integral analytically and we find
\begin{widetext}
\begin{equation}
\begin{aligned}
V_{CP}(x,y,z) = \frac{C_3}{2} &\left\{\frac{1}{x^3} + \frac{1}{(L-x)^3} + \frac{x^2 (z - \frac{W}{2} - x \tan \beta)^2 / 2 + \left( \frac{x}{\cos \beta} \right)^4}{(z - \frac{W}{2} - x \tan \beta)^3 x^3 \sqrt{x^2 + \left( z - \frac{W}{2} \right)^2}} \right. \\
&+ \frac{(z - \frac{W}{2} - x \tan \beta)^4 + (z - \frac{W}{2} - x \tan \beta) x \tan \beta \left[ (z - \frac{W}{2} - x \tan \beta)^2 + \left( \frac{x}{\cos \beta} \right)^2 \right]}{(z - \frac{W}{2} - x \tan \beta)^3 x^3 \sqrt{x^2 + \left( z - \frac{W}{2} \right)^2}} \\
&+ \frac{(z - \frac{W}{2} - x \tan \beta)^4 - (z - \frac{W}{2} - x \tan \beta) (L-x) \tan \beta \left[ (z - \frac{W}{2} - x \tan \beta)^2 + \left( \frac{L-x}{\cos \beta} \right)^2 \right]}{(z - \frac{W}{2} - x \tan \beta)^3 (L-x)^3 \sqrt{ \left( \frac{L-x}{\cos \beta} \right)^2 + \left( z - \frac{W}{2} -x\tan(\beta)\right)^2 -2(z- \frac{W}{2}-x\tan\beta)(L-x)\tan(\beta)}} \\
&+ \left.\frac{(L-x)^2 (z - \frac{W}{2} - x \tan \beta)^2 / 2 + \left( \frac{L-x}{\cos \beta} \right)^4}{(z - \frac{W}{2} - x \tan \beta)^3 (L-x)^3 \sqrt{\left( \frac{L-x}{\cos \beta} \right)^2 + \left( z - \frac{W}{2} -x\tan(\beta)\right)^2 -2(z- \frac{W}{2}-x\tan\beta)(L-x)\tan(\beta)}} \right\}
\label{eqn:trapezoidalPot}
\end{aligned}
\end{equation}
\end{widetext}
where \( C_3 = \frac{\rho C_6 \pi}{6} \). The pairwise potential for the lower slit at $z=-W/2$ can be easily determined by substituting $z$ with $-z$ in the provided expression.

%%%%%%%%%%%%%%%%%%%%%%%%%%%%%%%%%%%%%%%%%%%%%%%%%%%%%%%%%%%%


%apsrev4-2.bst 2019-01-14 (MD) hand-edited version of apsrev4-1.bst
%Control: key (0)
%Control: author (8) initials jnrlst
%Control: editor formatted (1) identically to author
%Control: production of article title (0) allowed
%Control: page (0) single
%Control: year (1) truncated
%Control: production of eprint (0) enabled
\begin{thebibliography}{0}%
\makeatletter
\providecommand \@ifxundefined [1]{%
 \@ifx{#1\undefined}
}%
\providecommand \@ifnum [1]{%
 \ifnum #1\expandafter \@firstoftwo
 \else \expandafter \@secondoftwo
 \fi
}%
\providecommand \@ifx [1]{%
 \ifx #1\expandafter \@firstoftwo
 \else \expandafter \@secondoftwo
 \fi
}%
\providecommand \natexlab [1]{#1}%
\providecommand \enquote  [1]{``#1''}%
\providecommand \bibnamefont  [1]{#1}%
\providecommand \bibfnamefont [1]{#1}%
\providecommand \citenamefont [1]{#1}%
\providecommand \href@noop [0]{\@secondoftwo}%
\providecommand \href [0]{\begingroup \@sanitize@url \@href}%
\providecommand \@href[1]{\@@startlink{#1}\@@href}%
\providecommand \@@href[1]{\endgroup#1\@@endlink}%
\providecommand \@sanitize@url [0]{\catcode `\\12\catcode `\$12\catcode
  `\&12\catcode `\#12\catcode `\^12\catcode `\_12\catcode `\%12\relax}%
\providecommand \@@startlink[1]{}%
\providecommand \@@endlink[0]{}%
\providecommand \url  [0]{\begingroup\@sanitize@url \@url }%
\providecommand \@url [1]{\endgroup\@href {#1}{\urlprefix }}%
\providecommand \urlprefix  [0]{URL }%
\providecommand \Eprint [0]{\href }%
\providecommand \doibase [0]{https://doi.org/}%
\providecommand \selectlanguage [0]{\@gobble}%
\providecommand \bibinfo  [0]{\@secondoftwo}%
\providecommand \bibfield  [0]{\@secondoftwo}%
\providecommand \translation [1]{[#1]}%
\providecommand \BibitemOpen [0]{}%
\providecommand \bibitemStop [0]{}%
\providecommand \bibitemNoStop [0]{.\EOS\space}%
\providecommand \EOS [0]{\spacefactor3000\relax}%
\providecommand \BibitemShut  [1]{\csname bibitem#1\endcsname}%
\let\auto@bib@innerbib\@empty
%</preamble>
\end{thebibliography}%


\begin{thebibliography}{99}
\bibitem{Buhmann2012} S. Y. Buhmann, \textit{Dispersion Forces I: Macroscopic quantum electrodynamics and ground-state Casimir, Casimir–Polder and van der Waals forces} (Springer, Heidelberg, 2012).

\bibitem{Mitsch2014} R. Mitsch, C. Sayrin, B. Albrecht, P. Schneeweiss, and A. Rauschenbeutel, Nat. Commun. \textbf{5}, 5713 (2014).

\bibitem{Vetsch2010} E. Vetsch, D. Reitz, G. Sagué, R. Schmidt, S. T. Dawkins, and A. Rauschenbeutel, Phys. Rev. Lett. \textbf{104}0, 203603 (2010).

\bibitem{Deasy2014} K. Deasy, C. F. Phelan, V. G. Truong, S. Nic Chormaic, and M. Daly, New J. Phys. \textbf{16}, 053052 (2014).

\bibitem{Patterson2018} B. D. Patterson, P. Solano, P. S. Julienne, L. A. Orozco, and S. L. Rolston, Phys. Rev. A \textbf{97}, 032509 (2018).

\bibitem{Peyrot2019} T. Peyrot, N. Šibalić, Y. R. P. Sortais, A. Browaeys, A. Sargsyan, D. Sarkisyan, I. G. Hughes, and C. S. Adams, Phys. Rev. A \textbf{100}, 022503 (2019).

\bibitem{Peyrot2019_v2} T. Peyrot, C. Beurthe, S. Coumar, M. Roulliay, K. Perronet, P. Bonnay, C. S. Adams, A. Browaeys, and Y. R. P. Sortais, Opt. Lett. \textbf{44}, 1940 (2019).

\bibitem{Ritter2018} R. Ritter, N. Gruhler, H. Dobbertin, H. Kübler, S. Scheel, W. Pernice, T. Pfau, and R. Löw, Phys. Rev. X \textbf{8}, 021032 (2018).

\bibitem{Skljarow2022} A. Skljarow, H. Kübler, C. S. Adams, T. Pfau, R. Löw, and H. Alaeian, Phys. Rev. Research \textbf{4}, 023073 (2022).

\bibitem{Knappe2004} S. Knappe, V. Shah, P. D. D. Schwindt, L. Hollberg, J. Kitching, L.-A. Liew, and J. Moreland, Appl. Phys. Lett. \textbf{85}, 1460 (2004).

\bibitem{Nshii2013} C. C. Nshii, M. Vangeleyn, J. P. Cotter, P. F. Griffin, E. A. Hinds, C. N. Ironside, P. See, A. G. Sinclair, E. Riis and A. S. Arnold, Nature Nanotech \textbf{8}, 321-324 (2013).

\bibitem{Amico2017} L. Amico, G. Birkl, M. Boshier and L.C. Kwek, New J. Phys. 19 020201 (2017).  

\bibitem{Onofrio2006} R. Onofrio, New J. Phys. \textbf{8} 237 (2006). 

\bibitem{Decca2007} R. S. Decca, D. López, E. Fischbach, G. L. Klimchitskaya, D. E. Krause, and V. M. Mostepanenko, Phys. Rev. D \textbf{75}, 077101 (2007).

\bibitem{Laliotis2021} A. Laliotis, B. Lu, M. Ducloy and D. Wilkowski, AVS Quantum Sci. \textbf{3}, 043501 (2021).

\bibitem{Bender2014} H. Bender, C. Stehle, C. Zimmermann, S. Slama, J. Fiedler, S. Scheel, S. Yoshi Buhmann, and V. N. Marachevsky, Phys. Rev. X \textbf{4}, 011029 (2014). 

\bibitem{Balland2023} Y. Balland, L. Absil and F. Pereira Dos Santos, Quectonewton local force sensor, arXiv:2310.14717 (2023).

\bibitem{Whittaker2014} K.A. Whittaker, J. Keaveney, I.G. Hughes, A. Sargsyan, D. Sarkisyan, and C.S. Adams, Optical Response of Gas-Phase Atoms at Less than $\lambda$/80 from a Dielectric Surface, Phys. Rev. Lett. \textbf{112}, 253201 (2014).

\bibitem{Carvalho2018} J. C. de Aquino Carvalho, P. Pedri, M. Ducloy, and A. Laliotis, Phys. Rev. A \textbf{97}, 023806 (2018). 

\bibitem{Lonij2009} V.t P. A. Lonij, W. F. Holmgren, and A. D. Cronin, Phys. Rev. A \textbf{80}, 062904 (2009).  

\bibitem{Garcion2021} C. Garcion, N. Fabre, H. Bricha, F. Perales, S. Scheel, M. Ducloy, and G. Dutier, Intermediate-Range Casimir-Polder Interaction Probed by High-Order Slow Atom Diffraction, Phys. Rev. Lett. \textbf{127}, 170402 (2021).

\bibitem{Morley2021} J Morley, R. Flack, B. J. Hiley, P. F. Barker, J. Phys. B: At. Mol. Opt. Phys. \textbf{54} 155301 (2021). 

\bibitem{Garcion2024} C. Garcion, Q. Bouton, J. Lecoffre, Nathalie Fabre, E. Charron, G. Dutier, and N. Gaaloul, Quantum description of atomic diffraction by material nanostructures, Phys. Rev. Res. \textbf{6}, 023165 (2024).

\bibitem{Buhmann2004} S Y Buhmann, Ho Trung Dung, and D-G Welsch., The van der waals energy of atomic systems near absorbing and dispersing bodies. Journal of Optics B: Quantum and Semiclassical Optics, \textbf{6}, S127–S135 (2004).

\bibitem{Scheel2008} S. Scheel and S.Y. Buhmann, Macroscopic quantum electrodynamics - Concepts and applications, Act. Phys. Slov., \textbf{58}, 675 – 809 (2008).

\bibitem{Perreault2005} J. D. Perreault, A. D. Cronin, and T. A. Savas, Using atomic diffraction of Na from material gratings to measure atom-surface interactions, Phys. Rev. A \textbf{71}, 053612 (2005).

\bibitem{Lepoutre2009} S. Lepoutre, H. Jelassi, V. P. A. Lonij, G. Tr´enec, M. Büchne1, A. D. Cronin and J. Vigué, Dispersive atom interferometry phase shifts due to atom-surface interactions, EPL, \textbf{88}, 20002 (2009).

\bibitem{Vassen2012} W. Vassen, C. Cohen-Tannoudji, M. Leduc, D. Boiron, C.I. Westbrook, A. Truscott, K. Baldwin, G. Birkl, P. Cancio, and M. Trippenbach, Cold and trapped metastable noble gases, Rev. Mod. Phys. \textbf{84}, 175 (2012).

\bibitem{Mostepanenko2016} V. M. Mostepanenko, Progress in constraining axion and non-Newtonian gravity from the Casimir effect, Int. J. Mod. phys. A \textbf{31}, 1641020 (2016).

\bibitem{baker_clarification_1984} S. Baker, R.D. Cousins, Clarification of the use of {CHI}-square and likelihood functions in fits to histograms, Nucl. Instrum. Methods Phys. Res. \textbf{2}, 221 (1984).

\bibitem{Fiedler2022} J. Fiedler, B. Holst, An atom passing through a hole in a dielectric membrane: impact of dispersion forces on mask-based matter-wave lithography, J. Phys. B: At. Mol. Opt. Phys. \textbf{55} 025401 (2022).

\bibitem{P_value_remark} The $P$-value is defined as $P = \int_{\chi_{min}^{2}}^{\infty} f(\chi_{red}^{2}) d\chi_{red}^{2} $, where $f(\chi_{red}^{2})$ is the $\chi_{red}^{2}$-distribution. If the distribution is Gaussian with a mean value of $\mu$ and a standard deviation $\sigma$,  we have $P= \frac{1}{2} - \frac{1}{2} \mathrm{erf} \left( \frac{\chi_{min}^{2}-\mu}{\sqrt{2} \sigma} \right)$.

\bibitem{Bruhl2002} R. Brühl, P. Fouquet, R. E. Grisenti, J. P. Toennies, G. C. Hegerfeldt, T. Köhler, M. Stoll and C. Walter, The van der Waals potential between metastable atoms and solid surfaces: Novel diffraction experiments vs. theory, EPL \textbf{59}, 357 (2002).

\bibitem{Hornberger2012} K. Hornberger, S. Gerlich, P. Haslinger, S. Nimmrichter, and M.s Arndt, Colloquium: Quantum interference of clusters and molecules, Rev. Mod. Phys. \textbf{84}, 157 (2012).

\bibitem{Grisenti2000} R. E. Grisenti, W. Schöllkopf, J. P. Toennies, J. R. Manson, T. A. Savas, and H. I. Smith, He-atom diffraction from nanostructure transmission gratings: The role of imperfections, Phys. Rev. A \textbf{61}, 033608 (2000).

\bibitem{Bruch2007} L. W. Bruch, R. D. Diehl, and J. A. Venables, Progress in the measurement and modeling of physisorbed layers, Rev. Mod. Phys. \textbf{79}, 1381 (2007).

\bibitem{Debiossac2014} M. Debiossac, A. Zugarramurdi, P. Lunca-Popa, A. Momeni, H. Khemliche, A.G Borisov, and P. Roncin, Transient Quantum Trapping of Fast Atoms at Surfaces, Phys. Rev. Lett. \textbf{112}, 023203 (2014).

\bibitem{Chang2014} D.E. Chang, K. Sinha, J.M. Taylor, H.J. Kimble, Trapping atoms using nanoscale quantum vacuum forces, Nat. Commun \textbf{5}, 4343 (2014).

\bibitem{Lifshitz1956} E. M. Lifshitz, The theory of molecular attractive forces between solids, Sov. Phys. JETP, \textbf{2}, 73 (1956).

\bibitem{Mavroyannis1963} C. Mavroyannis, The interaction of neutral molecules with dielectric surfaces, Mol. Phys. \textbf{6}, 593 (1963).

\bibitem{Karam2005} C Karam, N Wipf, J Grucker, F Perales, M Boustimi, G Vassilev,
V Bocvarski, C Mainos, J Baudon and J Robert, Atom diffraction with a ‘natural’ metastable atom
nozzle beam, J. Phys. B: At. Mol. Opt. Phys. \textbf{38}, 2691–2700 (2005).

\bibitem{Stielow2021} T. Stielow and S. Scheel, Reconstruction of nanoscale particles from single-shot wide-angle FEL diffractions patterns with physics-informed neural networks, Phys. Rev. E \textbf{103}, 053312 (2021).

\bibitem{Antoniadis2011} I. Antoniadis, S. Baessler, M. Büchner, V.V. Fedorove, S. Hoedl, A. Lambrecht, V.V. Nesvizhevsky, G. Pignol, K.V. Protasov, S. Reynaud, Yu. Sobolev, Short-range fundamental forces, C. R. Physique \textbf{12}, 755 (2011).

\bibitem{Philipp1973} H. R. Philipp, Optical properties of silicon
nitride, J. Electrochem. Soc. \textbf{120}, 295 (1973).

%%%%%%%%%%%%%%%%%%%%%%%%%%%%%%%%%%%%%%%%%%%%%%%%%%%%%%%%%%%%%%
%%APPENDIX%%

\bibitem{Luke2015} K. Luke, Y. Okawachi, M. R. E. Lamont, A. L. Gaeta, and M. Lipson, Broadband mid-infrared frequency comb generation in a $\mathrm{Si_{3}N_{4}}$ microresonator, Opt. Lett. \textbf{40}, 4823 (2015).

\bibitem{Fiedler2019} J. Fiedler, F. Spallek, P. Thiyam, C. Persson, M. Boström, M. Walter, and S. . Buhmann, Dispersion forces in inhomogeneous planarly layered media: A one-dimensional model for effective polarizabilities, Phys. Rev. A \textbf{99}, 062512 (2019).

\bibitem{Bitbol2013} A. F. Bitbol, A. Canaguier-Durand, A. Lambrecht, and S. Reynaud, Pairwise summation approximation for Casimir potentials and its limitations, Phys. Rev. B \textbf{87}, 045413 (2013).

\end{thebibliography}
\end{document}